\UseRawInputEncoding 
\documentclass[10pt,journal,compsoc]{IEEEtran}

\usepackage{fontawesome5}
\usepackage{graphicx}
\usepackage[x11names,table,dvipsnames,xcdraw]{xcolor}
\usepackage{booktabs}
\usepackage{multirow}
\usepackage[normalem]{ulem}
\useunder{\uline}{\ul}{}
\usepackage{subfig}
\usepackage{comment}
\usepackage{array, multirow, bigdelim, makecell}
\usepackage[skins]{tcolorbox} 
\usepackage{dblfloatfix}
\newcommand{\tabitem}{~~\llap{\textbullet}~~}
\usepackage{xurl}

\newcommand{\todo}[1]{}
\renewcommand{\todo}[1]{{\color{red} TODO: {#1}}}

\def\mybarhhigh#1#2{
   {\color{black}\rule{#1mm}{8pt}}  #2}
   
\def\mybar2#1#2#3#4#5{
   {\color{black}\rule{4pt}{#1mm}}
 {\color{black}\rule{4pt}{#2mm}}
  {\color{black}\rule{4pt}{#3mm}}
   {\color{black}\rule{4pt}{#4mm}}
    {\color{black}\rule{4pt}{#5mm}}
   }

\ifCLASSOPTIONcompsoc
  \usepackage[nocompress]{cite}
\else
  \usepackage{cite}
\fi
\ifCLASSINFOpdf
\else
\fi
\begin{document}

\title{The Emotional Roller Coaster of Responding to Requirements Changes in Software Engineering} 

\author{Kashumi Madampe,~\IEEEmembership{Graduate Student Member,~IEEE,}
        Rashina~Hoda,~\IEEEmembership{Member,~IEEE,}
        and~John~Grundy,~\IEEEmembership{Senior Member,~IEEE}
\IEEEcompsocitemizethanks{\IEEEcompsocthanksitem K. Madampe, R. Hoda, and J. Grundy are with the HumaniSE Lab at Department
of Software Systems and Cybersecurity, Faculty of Information Technology, Monash University, Wellington Road, Clayton, VIC 3800, Australia.\protect\\
E-mail: kashumi.madampe@monash.edu}
\thanks{Manuscript received September 08, 2021; revised Month Date, 2021.}}

\markboth{Submitted to IEEE Transactions on Software Engineering}%
{Shell \MakeLowercase{\textit{et al.}}: Bare Advanced Demo of IEEEtran.cls for IEEE Computer Society Journals}

\IEEEtitleabstractindextext{%
\begin{abstract}
\textbf{Background:} A preliminary study we conducted showed that software practitioners respond to requirements changes (RCs) with different emotions, and that their emotions vary at stages of the RC handling life cycle, such as \textit{receiving, developing,} and \textit{delivering} RCs.
\textbf{Objective:} We wanted to study more comprehensively how practitioners \textit{emotionally} respond to RCs. 
\textbf{Method:} We conducted a world-wide survey with the participation of 201 software practitioners. In our survey, we used the Job-related Affective Well-being Scale (JAWS) and open-ended questions to capture participants’ emotions when handling RCs in their work and query about the different circumstances when they feel these emotions. We used a combined approach of statistical analysis, JAWS, and Socio-Technical Grounded Theory (STGT) \textit{for Data Analysis} to analyse our survey data.
\textbf{Findings:} We identified 
(1) emotional responses to RCs, i.e., the most common \textit{emotions} felt by practitioners when handling RCs; (2) different \textit{stimuli} -- such as the RC, the practitioner, team, manager, customer -- that trigger these emotions through their own different characteristics; (3) \textit{emotion dynamics}, i.e., the changes in emotions during the project and RC handling life cycles; 
(4) \textit{distinct events} where particular emotions are triggered: project milestones, and RC stages; (5) and \textit{time related matters} that regulate the emotion dynamics.
\textbf{Conclusion:} 
Practitioners are not pleased with receiving RCs all the time. Last minute RCs introduced closer to a deadline especially violate emotional well-being of practitioners.
We present some practical recommendations for practitioners to follow, including a dual-purpose emotion-centric decision guide to help decide when to introduce or accept an RC, and some future key research directions.
\end{abstract}

\begin{IEEEkeywords}
emotions, affects, requirements, changes, human factors, mixed-methods, software engineering, software teams, socio-technical grounded theory 
\end{IEEEkeywords}}

\maketitle

\IEEEdisplaynontitleabstractindextext

\IEEEpeerreviewmaketitle

\section{Introduction}
\label{sec:introduction}
\IEEEPARstart{R}{equirements} changes (RCs) in software development include such actions as  additions, modifications and deletions of functional or non-functional requirements \cite{Madampe2021AContexts}. Such RCs naturally impact the cost, quality, and schedule of the project \cite{McGee2009ATaxonomy}. Hence, they are crucial to understand during software development. In traditional software development contexts, introducing an RC can be considered an intervention -- sometimes unexpected -- during the software development process. In agile contexts, requirements are welcomed ``\textit{even late in development}'' \cite{Beck2001ManifestoDevelopment}. Either way, an RC acts as a stimulus in the development environment with the potential to elicit responses from the practitioners handling the RC. Humans respond to stimuli in various ways. One such way is through \textit{emotions}. Emotions are defined as \textit{``a sequence of interrelated, synchronised changes in the states of all the five organismic subsystems (information processing, support, executive, action, and monitoring) in response to the evaluation of an external or internal stimulus event as relevant to central concerns of the organism''} \cite{Scherer1987TowardEmotion}, for example \textit{excitement, satisfaction, anxiety,} and \textit{fatigue} \cite{VanKatwyk2000UsingStressors.}, and reaction to a given stimulus through emotions is called \textit{emotional response} \cite{VandenBos2007APAPsychology}. Emotions act as behavioural motivators \cite{Colomo-Palacios2010AEngineering}, and have direct linkages to cognition [3], productivity \cite{Colomo-Palacios2010AEngineering},\cite{Graziotin2015DoEngineering},\cite{Kolakowska2013EmotionEngineering}, and decision-making \cite{Muller2015}. 

Research on understanding emotions of developers during software development contexts is gaining momentum. Several studies have focused on emotions and productivity of individuals in the software development teams \cite{Girardi2021EmotionsWorkplace}, \cite{Wrobel2016TowardsTeams}, \cite{Wrobel2013}, \cite{Crawford2014}, \cite{Graziotin2014}, \cite{Graziotin2014SoftwarePerformance}. However, only a few studies have focused on emotions of software development teams during requirements engineering (RE) \cite{Colomo-Palacios2010AEngineering, Colomo-Palacios2011UsingEngineering}. Through an interview-based preliminary study, we found that software development teams show emotional responses to RCs at stages of \textit{receiving}, \textit{developing}, and \textit{delivering} the RC \cite{Madampe2020TowardsTeams}. This preliminary study motivated us to conduct a more in-depth study on how emotions vary over the RC handling life cycle.  

Consider a software development team working \textit{``enthusiastically"}. Suddenly, the customer decides to change an already implemented requirement. Kash, who is a developer in the team, gets \textit{``angry"} because of this RC arriving, which distracts from current work, potentially wastes previous efforts, and goes against requests the customer previously made. Kash thinks that the customer does not understand Kash's job and its demands. At the same time, this RC makes her peers \textit{``angry''} and even \textit{``depressed"} about flow-on consequences of the new RC. The team disagree about certain things about handling the RC. Kash feels that her manager can not appreciate her team's emotions. However, she has no other option than working on the RC with her team. Therefore, she starts to implement the requested RC. At the beginning, Kash begins to feel more \textit{``energetic"} while updating the code, and once she is done with coding, she feels \textit{``inspired"} by solving the customer's new challenge. Then Kash decides to test the code, and during delivery she is \textit{``excited"}, and \textit{``inspired"}. Once the RC is delivered, Kash is fully \textit{``relaxed"}. Is the reality for practitioners similar or very different to this imaginary situation? What emotions do practitioners actually experience when responding to different kinds of RCs from different people at different times?  Is RC the only stimulus that triggers practitioners' emotions?  How do practitioners and their managers handle problematic emotions, and how do they encourage positive emotions?

To gain a more comprehensive understanding of emotional responses to RCs in software teams, our broad question is as follows:

 \begin{center}
         \textbf{How do software practitioners respond emotionally to requirement changes?}
    \end{center}
    

    


To answer this research question, we conducted a worldwide survey\footnote{Approved by Monash Human Research Ethics Committee. Approval Number: 23578} with the participation of 201 software practitioners. We utilised the Job-related Affective Well-being Scale (JAWS) \cite{VanKatwyk2000UsingStressors.} which assesses people's emotional reactions to their job during the past 30 days and a set of open-ended questions. We used a combined approach using descriptive statistical analysis, JAWS and socio-technical grounded theory (STGT) \cite{Hoda2021Socio-TechnicalEngineeringb} to analyse the data. 

In this paper, we use the terms ``emotional responses'' and ``feeling emotions/ respond with emotions/ experiencing emotions'' interchangeably. In simple terms, ``emotional response'' is bringing ``emotion'' into play. Our analysis resulted in identifying the following key findings:

\begin{itemize}
\item[KF1. ] \textbf{Emotional responses to RCs:} We found the most common emotions felt by practitioners when handling RCs;
\item[KF2. ]  \textbf{Stimuli triggering emotions:} We found several stimuli, such as the RC, the individual practitioner, the team, manager and customer, can all lead to the triggering of different emotions;
\item[KF3. ]  \textbf{Emotion dynamics in software project and requirements changes handling life cycles:} We found the phenomenon of emotion dynamics in software team contexts. i.e., the fluctuation of emotional responses across time during the project and the RC handling life cycles;
\item[KF4. ]  \textbf{Distinct events lead to emotion dynamics:} We found that practitioners' emotions are triggered at specific milestones of the project life cycle and stages of the RC handling life cycle;
\item[KF5. ]  \textbf{Regulation of emotion dynamics:} We discovered that temporal matters i.e., time related concerns have the capability of regulating the emotional responses of practitioners  at  project milestones and RC handling stages.

\end{itemize}





\noindent The key contributions of this research include:
\begin{itemize}
\item \textbf{Identification of a range of emotions and stimuli.} We identified and categorised a range of emotions and factors likely causing emotional responses to RCs, including when emotion occurs and other temporal matters, 
factors related to the individual, manager, team and customer. 

\item \textbf{A set of practical recommendations for practitioners} to follow in RC abundant environments;
\item \textbf{A dual-purpose emotion-centric decision guide for both carriers of RCs and practitioners who do not act as carriers of RCs} to use to decide when to introduce an RC to the team and when to accept an RC, respectively;
\item \textbf{Knowledge including a model representing an emerging theory} to learn and conduct further research about emotion dynamics in software team contexts. 
\end{itemize}

\section{Study Design}
\label{sec:mtd}

\subsection{Definitions}
We use some concepts from Psychology, Grounded Theory, and a few of our own terms throughout this paper. Table \ref{tab:definitions} presents the definitions of these terms. 
The cited definitions are directly from their sources and not paraphrased. We use the term ``individual cognition'' instead of the original term ``cognition'' to avoid the confusions with ``social cognition''. Similarly, ``team dynamics'', and ``team cohesion'' are used instead of ``group dynamics'' and ``group cohesion''.


\begin{table}[]
\caption{Definition of Key Terms Used}
\label{tab:definitions}
\scalebox{0.95}{
\resizebox{\columnwidth}{!}{%
\begin{tabular}{@{}ll@{}}

\toprule
\footnotesize
\textbf{Term}                                                       & \textbf{Definition}                                                                                                                                                                                                                                                      \\ \midrule
\footnotesize Individual cognition                                                & \begin{tabular}[c]{@{}l@{}}All forms of knowing and awareness, such as \\ perceiving, conceiving, remembering, reasoning, \\ judging, imagining, and  problem solving \cite{VandenBos2007APAPsychology}\end{tabular}                                                                                       \\ \cmidrule(r){1-2}
Cognitive skills                                                    & \begin{tabular}[c]{@{}l@{}}The skills involved in performing the tasks \\ associated with perception, learning, memory, \\ understanding, awareness, reasoning, judgement, \\ intuition, and language  \cite{VandenBos2007APAPsychology}\end{tabular}                                                       \\ \cmidrule(r){1-2}
Conation                                                            & \begin{tabular}[c]{@{}l@{}}The proactive (as opposed to habitual) part of \\ motivation that connects knowledge, affect, \\ drives, desires, and instincts to behavior  \cite{VandenBos2007APAPsychology}\end{tabular}                                                                                      \\ \cmidrule(r){1-2}
Emotion                                                             & \begin{tabular}[c]{@{}l@{}}a sequence of interrelated, synchronised changes \\in the states of all the five organismic subsystems \\(information processing, support, executive, \\action, and monitoring) in response to the evaluation \\of an external or internal stimulus event as relevant \\to central concerns of the organism \cite{Scherer1987TowardEmotion}\end{tabular}                                                           \\ \cmidrule(r){1-2}
Emotion dyamics                                                     & \begin{tabular}[c]{@{}l@{}}The patterns and regularities with which \\ emotions fluctuate over time \cite{Koval2015EmotionInertia}\end{tabular}                                                                                                                                                         \\ \cmidrule(r){1-2}
Emotional intelligence                                              & \begin{tabular}[c]{@{}l@{}}Type of intelligence that involves the ability to \\ process emotional information and use it in \\ reasoning and other cognitive activities  \cite{VandenBos2007APAPsychology}\end{tabular}                                                                                     \\ \cmidrule(r){1-2}
Emotion regulation                                                  & \begin{tabular}[c]{@{}l@{}}Any process that decreases, maintains, or increases \\ emotional intensity over time, thereby modifying the \\ spontaneous flow of emotions \cite{Koval2015EmotionInertia}, \cite{Gross2007EmotionPress.}, \cite{Koole2009TheReview}\end{tabular}                                                                                      \\ \cmidrule(r){1-2}
Emotional response                                                  & \begin{tabular}[c]{@{}l@{}}An emotional reaction, such as happiness, fear, or \\ sadness, to give a stimulus\  \cite{VandenBos2007APAPsychology}\end{tabular}                                                                                                                                                \\ \cmidrule(r){1-2}
Empathy                                                             & \begin{tabular}[c]{@{}l@{}}Understanding a person from his or her frame of \\ reference rather than one’s own, or vicariously\\ experiencing that person’s feelings, perceptions, \\ and thoughts  \cite{VandenBos2007APAPsychology}\end{tabular}                                                           \\ \cmidrule(r){1-2}
Motivation                                                          & \begin{tabular}[c]{@{}l@{}}A person’s willingness to exert physical or mental \\ effort in pursuit of a goal or outcome  \cite{VandenBos2007APAPsychology}\end{tabular}                                                                                                                                     \\ \cmidrule(r){1-2}
Perception                                                          & \begin{tabular}[c]{@{}l@{}}The process or result of becoming aware of objects, \\ relationships, and events by means of the senses, \\ which includes such activities as recognising, \\ observing, and discriminating  \cite{VandenBos2007APAPsychology}\end{tabular}                                      \\ \cmidrule(r){1-2}
Self-efficacy                                                       & \begin{tabular}[c]{@{}l@{}}An individual’s subjective perception of his or her \\ capability to perform in a given setting or to attain \\ desired results  \cite{VandenBos2007APAPsychology}\end{tabular}                                                                                                  \\ \cmidrule(r){1-2}
Social cognition                                                    & \begin{tabular}[c]{@{}l@{}}Cognition in which people perceive, think about, \\ interpret, categorise, and judge their own social \\ behaviors and those of others  \cite{VandenBos2007APAPsychology}\end{tabular}                                                                                           \\ \cmidrule(r){1-2}

Stimulus &
\begin{tabular}[c]{@{}l@{}}Any agent, event, or situation—internal or \\external—that elicits a response from an \\organism \cite{VandenBos2007APAPsychology}\end{tabular} \\
\cmidrule(r){1-2} 
Sustained attention                                                 & \begin{tabular}[c]{@{}l@{}}The ability to sustain attention over time in specific \\ goal-directed behaviors  \cite{Cristofori2015TraumaticCognition}\end{tabular}                                                                                                                                                \\ \cmidrule(r){1-2}
Team cohesion                                                       & \begin{tabular}[c]{@{}l@{}}The unity or solidarity of a group, including the \\ integration of the group for both social and \\ task-related purposes  \cite{VandenBos2007APAPsychology}\end{tabular}                                                                                                       \\ \cmidrule(r){1-2}
Team dynamics                                                       & \begin{tabular}[c]{@{}l@{}}The processes, operations, and changes that occur \\ within social groups, which affect patterns of \\ affiliation, communication, conflict, conformity, \\ decision making, influence, leadership, norm \\ formation, and power  \cite{VandenBos2007APAPsychology}\end{tabular} \\ \cmidrule(r){1-2}\morecmidrules\cmidrule(r){1-2}
Dimensions                                                          & \begin{tabular}[c]{@{}l@{}}The range along which general properties of a \\ category vary, giving specification to a category\\ and variation to a theory \cite{Strauss1998BasicsTechniques.}\end{tabular}                                                                                                   \\ \cmidrule(r){1-2}
Properties                                                           & \begin{tabular}[c]{@{}l@{}}Characteristics of a category, the delineation of \\ which defines and gives it meaning \cite{Strauss1998BasicsTechniques.}\end{tabular}                                                                                                                                          \\ \cmidrule(r){1-2}\morecmidrules\cmidrule(r){1-2}

Project                                                               & \begin{tabular}[c]{@{}l@{}}A temporary endeavor undertaken to create a \\ unique project service or result \cite{ProjectManagementInstitute2008PMBOKGUIDE}\end{tabular}                                                                                                                                                  \\ \cmidrule(r){1-2}
Project life cycle                                                    & \begin{tabular}[c]{@{}l@{}}The execution flow across time of any software \\ project that allows the software team to meet \\ the project goal\end{tabular}                                                                                                              \\ \cmidrule(r){1-2}
\begin{tabular}[c]{@{}l@{}}Requirements Change \\ (RC)   \end{tabular} & \begin{tabular}[c]{@{}l@{}}Additions/modifications/deletions of functional\\ /non-functional requirements in a software project\end{tabular}                                                                                                                             \\ \cmidrule(r){1-2}
RC handling life cycle                                                         & \begin{tabular}[c]{@{}l@{}}The execution flow across time of an RC that \\ allows the software team to deliver it. RC handling \\life cycle begins when the RC is introduced \\to the team and lies within the project life cycle\end{tabular}                                  \\ \cmidrule(r){1-2}
Team contexts                                                         & \begin{tabular}[c]{@{}l@{}}The socio-technical environment where software \\ practitioners work collectively as a team to achieve \\ a common goal\end{tabular}                                                                                                          \\ \cmidrule(r){1-2}
Temporal matters                                                      & \begin{tabular}[c]{@{}l@{}}Time related concerns that have impacts on project \\ and RC handling life cycles\end{tabular}   \\ \bottomrule
\end{tabular}}%
}
\end{table}

\subsection{Approach}
\begin{figure}
    \centering
    \includegraphics[width=\columnwidth]{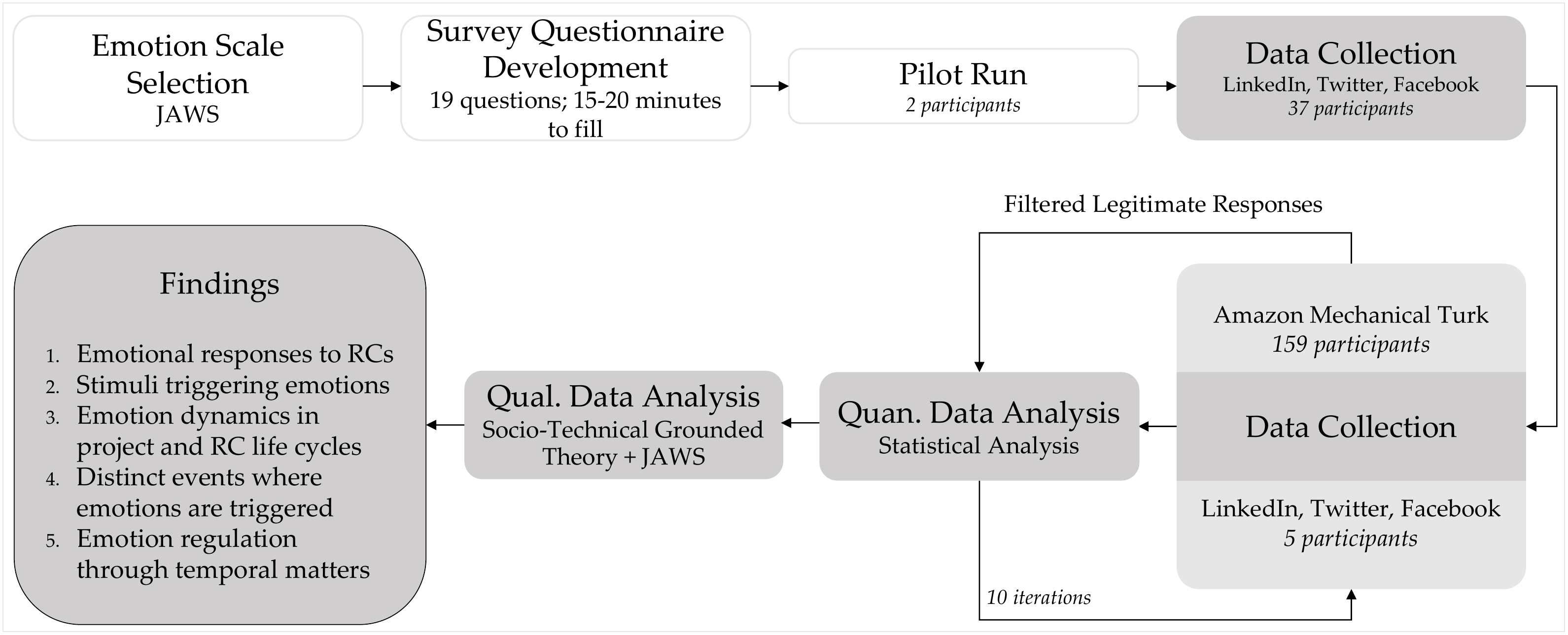}
    \caption{Study Approach (Quan: Quantitative; Qual: Qualitative; JAWS: Job-related Affective Well-being Scale)}
    \label{fig:mtd}
\end{figure}

An overview of our research approach is given in Fig. \ref{fig:mtd} and described in detail below. The replication package, including the survey questionnaire, demographic and project information of the participants, is available online\footnote{\url{https://github.com/kashumi-m/ReplicationPackageEmotionalRollerCoaster}}.

\subsubsection{[Step 1]: Emotion Scale Selection}
We need a way to describe human emotional responses to RCs. We evaluated 20 well-established emotion scales (15 as in \cite{Curumsing2017Emotion-orientedEngineering} and PANAS \cite{Watson1988DevelopmentScales.}, SPANE \cite{Diener2010NewFeelings}, JES \cite{Fisher1997}, DEQ \cite{UniversityofYork.NHSCentreforReviewsDissemination2009SystematicCare}, JAWS \cite{VanKatwyk2000UsingStressors.}) by comparing their listed emotions and their applicability to use to describe practitioners' emotional responses to RCs. From our analysis, we found 3 scales -- Discrete Emotions Questionnaire (DEQ), Job Emotion Scale (JES), and Job-related Affective Well-being Scale (JAWS) -- as suitable candidates. From our own industrial experience, we opted not to use DEQ as we found that some emotions were irrelevant for software development teams (e.g.: ``terror" and ``craving"). We had used JES in our previous work [11] which consists of 16 emotions. However, we wanted a comprehensive understanding of emotional responses to RCs. In the end we decided to use JAWS which has been used extensively to assess emotional reactions of people to their jobs over the past 30 days. As our survey questionnaire requested participants to respond by considering the current or most recent project they worked on, we found JAWS likely to be the best emotion scale for our study.

JAWS has two forms: one with 30 emotions (long form) and another with 20 emotions (short form). We used the short form which the authors of JAWS claim as the scale that is most commonly used \cite{Job-relatedSpector}.
The 20 emotions in JAWS are categorised into 4 sub-scales along the dimensions: pleasure and arousal (intensity). The sub-scales are namely, High pleasurable-High arousal (High\textsuperscript{2}), High pleasurable-Low Arousal (High\textsuperscript{1}), Low pleasurable-High Arousal (Low\textsuperscript{1}), and Low pleasurable-Low Arousal (Low\textsuperscript{2}). We abbreviated the sub-scales as above by making the abbreviation central to the pleasure. i.e., for example, when both pleasure and arousal are high, we abbreviated it as high\textsuperscript{2}; otherwise high\textsuperscript{1}. The emotions under each sub-scale are given in Table \ref{tab:JAWS}. The scale enables the participants to select one of the following five choices choice per emotion: \textit{never, rarely, sometimes, quite often,} and \textit{extremely often}. 

\begin{table}[]
\caption{Job-related Affective Well-being Scale Sub-Scales}
\label{tab:JAWS}
\resizebox{\columnwidth}{!}{%
\begin{tabular}{ll}
\toprule
\multicolumn{1}{c}{\textbf{Sub Scale}} & \multicolumn{1}{c}{\textbf{Emotion}}                 \\ \midrule
High\textsuperscript{2}                                   & Energetic, Excited, Ecstatic, Enthusiastic, Inspired \\
High\textsuperscript{1}                                   & At-ease, Calm, Content, Satisfied, Relaxed           \\
Low\textsuperscript{1}                                  & Angry, Anxious, Disgusted, Frightened, Furious       \\
Low\textsuperscript{2}                                   & Bored, Depressed, Discouraged, Gloomy, Fatigued      \\ \bottomrule
\end{tabular}%
}
\end{table}

\subsubsection{[Step 2]: Survey Questionnaire Development}
\begin{figure}
    \centering
    \includegraphics[width=\columnwidth]{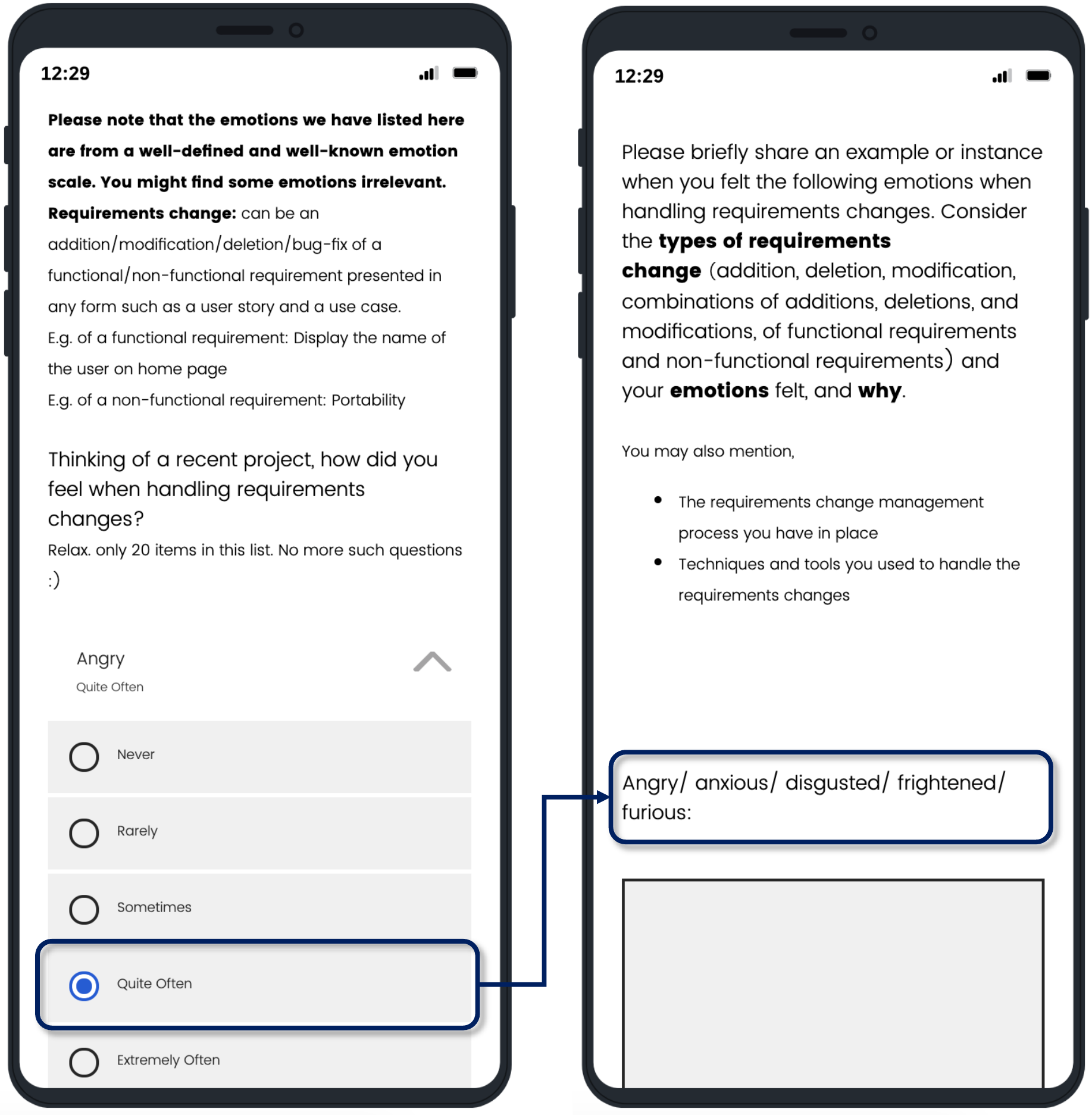}
    \caption{Prompting of Open-ended Questions based on Answers Given to the JAWS based Close-ended Question \small{(the selected choice shown here is for illustration purpose only, the data is not real)}}
    \label{fig:survey_questions}
\end{figure}

After we chose the emotion scale for our study, we developed the survey questionnaire by following Kitchenham et al.’s \cite{Kitchenham2008PersonalSurveys}, \cite{Kitchenham2002PreliminaryEngineering}, and Punter et al.’s\cite{Punter2003ConductingEngineering} guidelines. The survey consisted of 4 sections (demographics information, project information, team information, emotional responses to RCs). 15 closed-ended questions were distributed among these sections, and 4 open-ended questions belonged to the emotional responses to RCs section which was the JAWS scale. The 4 open-ended questions represented each sub-scale of JAWS. The open-ended questions were customised and prompted for participants based on the answers they gave to the closed-ended question on emotions felt when handling RCs. If the participant chose \textit{sometimes}, \textit{quite often}, or \textit{extremely often} for a particular emotion, the respective open-ended question was shown after answering the closed-ended question. This is illustrated in Fig. \ref{fig:survey_questions}, also showing samples of closed-ended and open-ended questions. 
In the example given in Fig. \ref{fig:survey_questions}, the choice ``quite often'' selected for the emotion \textit{angry} prompted the respective open-ended question.

We used \textit{Qualtrics}\footnote{https://www.qualtrics.com/}  as the survey platform and  distributed the survey  online following Smith et al.’s study \cite{Smith2013ImprovingSurveys} to improve the survey distribution. We did not collect any personal information from participants, except from the participants who voluntarily provided their details for future interviews. 

\subsubsection{[Step 3]: Data Collection}
Once the survey questionnaire was finalised, we sent the survey to 2 Ph.D. students who had recent industrial experience. They provided feedback about the survey in terms of time for completion. 
Before distributing the survey to the software development community, we calculated the sample size (random sampling) required for our study.
According to Global Developer Population Report \cite{SlashDataLtd.2019The2019}, the active developer population by 2019 was 18.9 M. We calculated the sample size required to generalise our findings to this population of 18.9 M. The sample size resulted in was N=384, by setting the confidence level to 95\% (Z-score=1.96), standard deviation to 0.5, and margin of error to 0.05. 

We distributed the survey by posting its link on social media such as \textit{LinkedIn, Twitter,} and \textit{Facebook}. The survey was available to the public for a period of 1 month and 20 days. The number of responses we received for the survey through this approach was limited. As we reached a plateau of 37 responses, we decided to recruit the rest of the participants via \textit{Amazon Mechanical Turk} (AMT)\footnote{https://www.mturk.com/}. The principal qualification criteria we used at AMT were:
   (1) \textit{Employment Industry -- Software and IT Services}; and
    (2) \textit{Job Function – Information Technology}. 
However, since we recruited the participants iteratively, we added additional criteria to different batches to have a uniform geographical and gender distribution as much as possible (purposive sampling). At the point of recruiting participants through AMT, we made a few changes to the survey questionnaire. 
One such change was making the open-ended questions mandatory. Later this became an advantage to us as we were able to filter the genuine responses by looking at the answers given to the open-ended questions. 49 responses were found as deceptive since unrelated answers were given to the open-ended questions. We rejected these responses and re-recruited participants until all responses appeared to be legitimate.
We initially targeted recruiting the calculated sample of 384 participants. However, as we analysed the quantitative data after each round of data collection, we stopped collecting data when we were able to see clear results. i.e., when the number of responses for each emotion in the emotion scale became steady. Thus, we ended up recruiting only 201 participants. The survey was not limited to collecting qualitative data, but also collected quantitative data. 

\begin{table}[]
\caption{\small Demographic Information of Participants (Dev: Developer; AC/SM: Agile Coach/Scrum Master; BA: Business Analyst; Other: $\leq$ 5 participants; XT: Total Software Development Experience; XTA: Total Agile Experience)}
\label{tab:demo_condensed}
\resizebox{\columnwidth}{!}{%
\begin{tabular}{@{}llll@{}}
\toprule
\textbf{Location} & \textbf{\# of Participants} & \textbf{Role} & \textbf{\# of Participants} \\ \midrule
North America     & \mybarhhigh{19.2}{96}                          & Dev           & \mybarhhigh{15}{75}                          \\
Asia              & \mybarhhigh{8}{40}                          & Manager       & \mybarhhigh{4.2}{21}                          \\
Europe            & \mybarhhigh{4.8}{24}                          & BA            & \mybarhhigh{3.8}{19}                          \\
Australasia       & \mybarhhigh{4.4}{22}                          & Dev, Tester   & \mybarhhigh{2.8}{14}                          \\
South America     & \mybarhhigh{3.4}{17}                          & Tester        & \mybarhhigh{2}{10}                          \\
Africa            & \mybarhhigh{0.4}{2}                           & Dev, Manager  & \mybarhhigh{1.8}{9}                           \\ \cmidrule(r){1-2}
\textbf{Gender}   & \textbf{\# of Participants} & AC/SM         & \mybarhhigh{1.6}{8}                           \\ \cmidrule(r){1-2}
Male              & \mybarhhigh{23}{115}                         & AC/SM, Dev    & \mybarhhigh{1.4}{7}                           \\
Female            & \mybarhhigh{17}{85}                        & Other         & \mybarhhigh{7.6}{38}                          \\
Gender diverse    & \mybarhhigh{0.2}{1}                           &               &                             \\ \midrule
\textbf{XT}       & \textbf{\# of Years}        & \textbf{XTA}  & \textbf{\# of Years}        \\ \midrule
Minimum           & 1                           & Minimum       & 0                           \\
Maximum           & 35                          & Maximum       & 20                          \\
Mean              & 7.84                        & Mean          & 5.12                        \\ \bottomrule
\end{tabular}%
}
\end{table}

\begin{table*}[]
\caption{\small Information of Current/Most Recent Project of the Participants (XP: Extreme Programming; Other: $\leq$ 5 participants)}
\label{tab:project_info_condensed}
\resizebox{\textwidth}{!}{%
\begin{tabular}{@{}llllll@{}}
\toprule
\footnotesize
\textbf{Project Domain} & \textbf{\# of Participants} & \textbf{Project Category} & \textbf{\# of Participants} & \textbf{Development Method Used} & \textbf{\# of Participants} \\ \midrule
IT                         & \mybarhhigh{12.2}{122}                         & New development           & \mybarhhigh{11.5}{115}                         & Scrum                            & \mybarhhigh{5.7}{57}                          \\
Finance \& Banking         & \mybarhhigh{3}{30}                          & Software as a Service     & \mybarhhigh{4.7}{47}                          & Dynamic System Development       & \mybarhhigh{4.4}{44}                          \\
Manufacturing              & \mybarhhigh{1}{10}                          & Maintenance               & \mybarhhigh{2.2}{22}                          & Feature Driven Development       & \mybarhhigh{2.5}{25}                          \\
Transport                  & \mybarhhigh{1}{10}                          & Migration                 & \mybarhhigh{1.7}{17}                          & Waterfall                        & \mybarhhigh{1.7}{17}                          \\
Telecom                    & \mybarhhigh{0.7}{7}                           &                           &                             & Kanban                           & \mybarhhigh{1.4}{14}                        \\
Other                   & \mybarhhigh{1.6}{16}                          &                           &                             & Crystal                          & \mybarhhigh{1.1}{11}                         \\ \cmidrule(r){1-4}
\textbf{Team Size }              & \textbf{\# of People}                & I\textbf{teration Length }         & \textbf{\# of Weeks}                 & None                             & \mybarhhigh{0.8}{8}                            \\ \cmidrule(r){1-4}
Minimum                 & 1                           & Minimum                   & 1                           & Other                            & \mybarhhigh{1.1}{11}                          \\
Maximum                 & 100                         & Maximum                   & 10                          &                                  &                             \\
Mean                    & 18.61                       & Mean                      & 5.67                        &                                  &                             \\
Standard Deviation      & 18.8                        & Standard Deviation        & 2.11                        &                                  &                                            \\ \midrule
\multicolumn{6}{l}{\textbf{Practices Followed   (Order of the Bars in Each Graph Below: Never $\rightarrow$ Sometimes $\rightarrow$ About half the time $\rightarrow$  Most of the time $\rightarrow$ Always)}}                            \\ \midrule
Collective Estimation      & \mybar2{0.3}{1.566666667}{1.366666667}{2.633333333}{0.833333333}               & Product Backlog           & \mybar2{0.333333333333333}{1.16666666666667}{1.33333333333333}{2.33333333333333}{1.53333333333333}              & Scrum/Kanban Board               & \mybar2{0.5}{1.03333333333333}{1.43333333333333}{1.96666666666667}{1.76666666666667}              \\
Customer Demos             & \mybar2{0.366666666666667}{1.23333333333333}{1.86666666666667}{2.1}{1.13333333333333}              & Short Iterations/Sprints  & \mybar2{0.333333333333333}{1.26666666666667}{1.56666666666667}{1.96666666666667}{1.56666666666667}              & Self-assignment                  & \mybar2{0.433333333333333}{1.16666666666667}{1.4}{2.3}{1.4}             \\
Daily Standup/team meeting & \mybar2{0.133333333333333}{0.933333333333333}{1.26666666666667}{1.86666666666667}{2.5}               & Release Planning          & \mybar2{0.2}{1.03333333333333}{1.43333333333333}{2.63333333333333}{1.4}               & Sprint Backlog                   & \mybar2{0.466666666666667}{1.5}{1.33333333333333}{2.06666666666667}{1.33333333333333}              \\
Definition of Done         & \mybar2{0.4}{1}{1.53333333333333}{2.33333333333333}{1.43333333333333}              & Retrospectives            & \mybar2{0.733333333333333}{1.4}{1.36666666666667}{1.83333333333333}{1.36666666666667}              & User Stories                     & \mybar2{0.533333333333333}{0.833333333333333}{1.56666666666667}{1.9}{1.86666666666667}              \\
Iteration Planning         & \mybar2{0.266666666666667}{0.833333333333333}{1.56666666666667}{2.6}{1.43333333333333}               & Review Meetings           & \mybar2{0.4}{1.06666666666667}{1.23333333333333}{1.93333333333333}{2.06666666666667}              & Use Cases                        & \mybar2{0.4}{0.8}{1.3}{2.53333333333333}{1.66666666666667}             \\
Pair Programming           & \mybar2{1.13333333333333}{1.5}{1.53333333333333}{1.6}{0.933333333333333}              &                           &                             &                                  &                             \\ \bottomrule        
\end{tabular}%
}
\end{table*}

The summary of demographic data of the participants is given in Table \ref{tab:demo_condensed}.
The majority of participants represented North America (N=96; 47.78\%) and the majority were developers (N=75; 37.31\%). The participants had a mean total experience of 7.8 years (min(total experience)=1 year; max(total experience)=35 years).
The summary of participants' most recent/current project is given in Table \ref{tab:project_info_condensed}. The answers they gave to the survey questionnaire were based on these projects. The majority of the participants' projects were new developments (N=115; 57.21\%), and the majority used agile methods in their projects (N=176; 87.56\%) which is in line with reported agile use in the industry \cite{202014thAgile}.

\subsubsection{[Step 4]: Data Analysis}
Table \ref{tab:mixed-methods} summarises how the data collected were analysed using a mixed-methods approach which included a quantitative analysis and a combined qualitative analysis. Our mixed-methods approach included using the same instrument (JAWS) for data collection and analysis, besides open-ended questions. JAWS served as a tool for collecting the emotional responses of the participants (resulting in KF1) through a close-ended matrix question and to analyse qualitative data (resulting in KF3) in a combined manner along with STGT (resulting in KF2, KF4, KF5). We describe the analysis process in detail below.

\begin{table}[]
\caption{\small Mixed-methods Approach (DC: Data Collection; DA: Data Analysis; JAWS: Job Affective Well-being Scale; STGT4DA: Socio-Technical Grounded Theory for Data Analysis; DA1: Combined Qualitative Data Analysis Step 1; DA2: Combined Qualitative Data Analysis Step 2)}
\label{tab:mixed-methods}
\resizebox{\columnwidth}{!}{%
\begin{tabular}{@{}llll@{}}
\toprule
\textbf{Method}                                                                 & \textbf{DC Technique}                                                            & \textbf{DA Technique}                                                      & \textbf{Key Finding}                                                     \\ \midrule
Quantitative                                                                    & \begin{tabular}[c]{@{}l@{}}Close-ended matrix :\\ JAWS Scale\end{tabular}        & \begin{tabular}[c]{@{}l@{}}Descriptive statistical\\ analysis\end{tabular} & KF1                                                                      \\ \midrule
\multirow{2}{*}{\begin{tabular}[c]{@{}l@{}}Combined\\ qualitative\end{tabular}} & \multirow{2}{*}{\begin{tabular}[c]{@{}l@{}}Open-ended \\ questions\end{tabular}} & \begin{tabular}[c]{@{}l@{}}STGT4DA+JAWS \\ (DA1)\end{tabular}              & \begin{tabular}[c]{@{}l@{}}KF2 (emerging), \\ KF3, KF4, KF5\end{tabular} \\ \cmidrule(l){3-4} 
                                                                                &                                                                                  & STGT4DA (DA2)                                                              & KF2 (emerged)                                                            \\ \bottomrule
\end{tabular}%
}
\end{table}

\textbf{Quantitative Analysis:}
The data collected through JAWS were descriptively analysed using \textit{Python}. This resulted in emotional responses to RCs and Table \ref{tab:all_emotions} presents the analysis results.


\textbf{Combined Qualitative Analysis:}
The collected qualitative data was stored and analysed in \textit{MAXQDA}\footnote{https://www.maxqda.com/}. Additionally, \textit{Microsoft Excel} was used when necessary. Categories of stimuli, emotions, emotion dynamics, distinct events, and temporal matters were identified from a combined approach of JAWS and \textit{STGT for data analysis}
\cite{Hoda2021Socio-TechnicalEngineeringb}. We decided to use \textit{STGT for data analysis} due to its ability to apply the basic data analysis techniques, such as open coding, constant comparison, and memoing, within mixed-methods studies, in a limited capacity (as opposed to full theory development) and its suitability to analyse data in socio-technical research contexts. Our previous experience with applying open coding and constant comparison using the Strauss-Corbinian GT \cite{Strauss1998BasicsTechniques.} also helped.

\begin{figure}[b]
    \centering
    \includegraphics[width=\columnwidth]{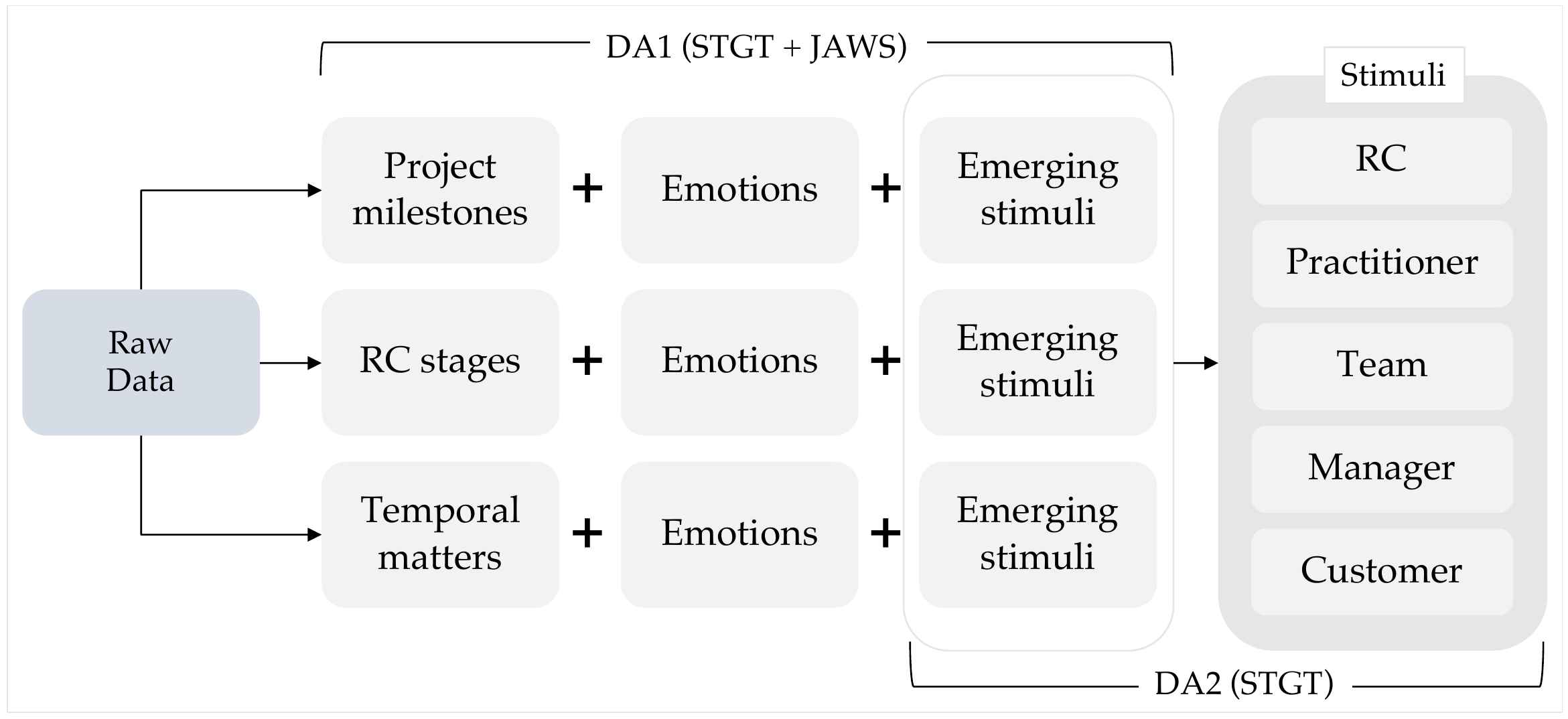}
    \caption{Combined Qualitative Data Analysis Approach (STGT: Socio-Technical Grounded Theory; JAWS: Job-related Affective Well-being Scale)}
    \label{fig:data_analysis}
\end{figure}

The combined qualitative analysis occurred in two steps: Combined Qualitative Data Analysis Step 1 (DA1) and Combined Qualitative Data Analysis Step 2 (DA2), as shown in Fig. \ref{fig:data_analysis}. The analysis of raw data in DA1 yielded concepts such as \textit{project milestones, RC stages,} and \textit{temporal matters} along with their respective triggered \textit{emotions}, and emerging \textit{stimuli} that triggered them. Further analysis of emerging stimuli found in DA1, allowed us to group them under \textit{RC, practitioner, team, manager,} and \textit{customer}, in DA2. DA1 and DA2 are explained in detail below.



\textbf{DA1 (STGT and JAWS):} DA1 consisted of a combination of STGT basic data analysis and JAWS. STGT analysis resulted in identifying project milestones, RC stages, and emerging stimuli and JAWS helped in identifying emotions. 

\textit{STGT data analysis:} We followed a micro-analysis (line-by-line) approach in identifying the fragments of data in textual open-ended answers by the participants, since survey responses were 1--3 lines each. These identified data fragments were then meaningfully labelled (as codes). Through constant comparison, we grouped similar codes to form concepts, and then grouped similar concepts to form sub-categories, which are explainable through their characteristics (properties and dimensions). After the sub-category development, we further applied constant comparison to form categories.

\textit{JAWS data analysis:} Since the open-ended questions listed the emotions in JAWS, most participants used the exact terms of emotions as in JAWS (see Fig. \ref{fig:survey_questions}). When the answers in the open-text did not list any specific emotion, we considered the participant felt the majority of emotions in the emotion sub-scale of the question. This allowed us to extract the emotions from participants’ answers to the open-ended questions. 

\begin{figure*}
    \centering
    \includegraphics[width=\textwidth]{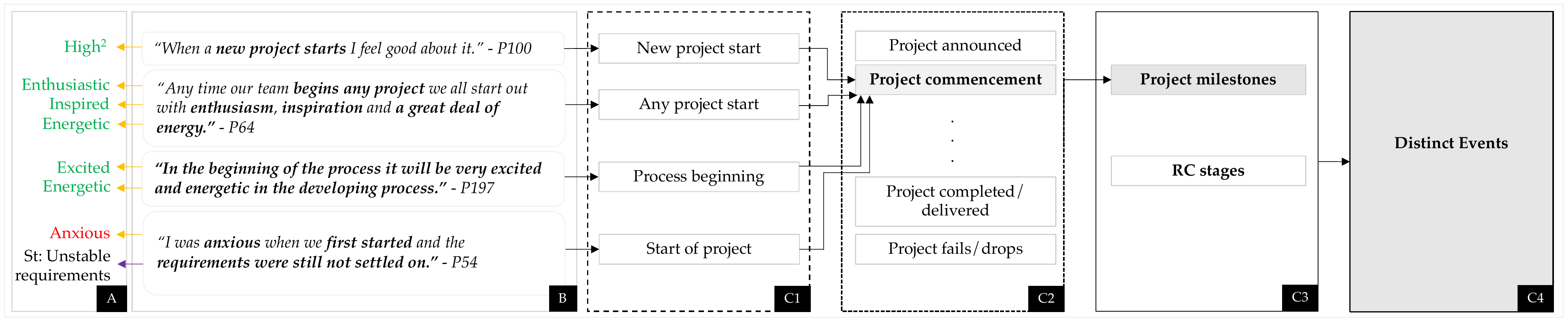}
    \caption{Combined Qualitative Data Analysis Step 1 Example: Emergence of the Category ``Distinct Events''; Capturing of Emotions, and Stimuli (St: Stimulus), using JAWS (Block A) and STGT for Data Analysis (right-hand side of Block B)}
    \label{fig:da1_example}
\end{figure*}

The extreme left-hand side of Fig. \ref{fig:da1_example} (block A) shows the extraction of emotions and emerging stimuli at code level from the raw data (block B). Blocks to the right of the raw data show the emergence of the category “distinct events” in the order of analysis of raw data (block B) $\rightarrow$ codes (C1) $\rightarrow$ concepts (C2) $\rightarrow$ sub-categories (C3) $\rightarrow$ category (C4). Below, we elaborate this further by taking a single data point from Fig. \ref{fig:da1_example}. This data point represents the raw data that consists of (a) the distinct event at granular level, i.e., the code -- ``start of project’’, which led to the concept -- ``project commencement’’ with other similar codes as in Fig. \ref{fig:da1_example}, that later resulted in the sub-category -- ``project milestone’’ and the category -- “distinct events”, (b) emerging stimulus which is at code level -- ``unstable requirements’’, and (c) the emotion -- ``anxious’’.



\begin{tcolorbox}[colback=gray!10, boxrule=0pt,frame hidden]
\textbf{Raw data:} \textit{``I was anxious when we first started and the requirements were still not settled on'' -- P54 [Tester]}\\

\textbf{[JAWS] Emotion:} Anxious\\

\textbf{[STGT] Code:} Start of project\\
\textbf{[STGT] Concept:} Project commencement\\
\textbf{[STGT] Sub-category:} Project milestone\\
\textbf{[STGT] Category:} Distinct events\\

\textbf{[STGT] Code (leading to the emergence of ``RC as a stimulus'' in DA2):} Unstable requirements
\end{tcolorbox}

\textbf{DA2 (STGT):} 
We further analysed the emerging stimuli that we found during DA1. We identified that these are certain aspects of \textit{RC, practitioners, team, manager,} and  \textit{customer}. Grouping these together led to the category ``stimuli''.

During DA2, we also identified that a set of codes belong to certain well-established principles in Psychology (see Table \ref{tab:definitions}). Therefore, we used those principles to conceptualise, and name the concepts ``individual conation'', ``individual cognition'', ``social cognition'', ``team dynamics'', and ``emotional intelligence''. In the  example below, mapping to these Psychological concepts, we show the further analysis of emerging stimuli that we found during DA1.



\begin{tcolorbox}[colback=gray!10, boxrule=0pt, frame hidden]
\textbf{[STGT] Code (captured during DA1):} Practitioner's sustained attention\\
\textbf{[STGT] Property [$<$dimension$>$]:} Self cognitive skill [+]\\
\textbf{[STGT] Concept:} Individual cognition\\
\textbf{[STGT] Sub-category:} Practitioner as a stimulus\\
\textbf{[STGT] Category:} Stimuli
\end{tcolorbox}

Further examples of DA2 which applied STGT for data analysis is given in Table \ref{tab:DA2_example}.

\begin{table*}
\caption{Combined Qualitative Data Analysis Step 2 Example (Emergence of Sub-Category: ``Practitioner as a Stimulus'' from Stimuli found during Data Analysis Step 1)}
\label{tab:DA2_example}
\resizebox{\textwidth}{!}{%
\begin{tabular}{@{}llll@{}}
\toprule
\footnotesize
\textbf{Code (Emerging stimuli after DA1)}                                                                                                   & \textbf{Property {[}\textless{}dimension\textgreater{}{]}}                   & \textbf{Concept}                      & \textbf{Sub-category (Stimuli after DA2)}    \\ \midrule
\tabitem Practitioner motivated to put much effort                                                                                           & \hspace{-1em}\rdelim\}{2}{*}[Self motivation {[}+{]}]                                     & \hspace{-1em}\rdelim\}{2.8}{*}[Individual conation]  & \hspace{-1em}\rdelim\}{20}{*}[Practitioner as a stimulus] \\
\tabitem Practitioner motivated as not coded in a while                                                                                      &                                                                              &                                       &                                              \\
\tabitem Practitioner feeling that hard work is wasted                                                                                       & \hspace{-1em}\rdelim\}{1}{*}[Self motivation {[}-{]}]                                                      &                                       &                                              \\ \cmidrule(r){1-3}
\begin{tabular}[c]{@{}l@{}}\tabitem Practitioner knowing the possibility of \\ successful delivery\end{tabular}                              & \hspace{-1em}\rdelim\}{1.2}{*}[Self-efficacy {[}+{]}]                                                        & \hspace{-1em}\rdelim\}{6.5}{*}[Individual cognition] &                                              \\
\begin{tabular}[c]{@{}l@{}}\tabitem Practitioner perceiving him/her/themselves\\  as a perfectionist\end{tabular}                            & \hspace{-1em}\rdelim\}{1}{*}[Personality perception [self]]  &                                       &                                              \\
\tabitem Practitioner's sustained attention                                                                                                  & \hspace{-1em}\rdelim\}{2.5}{*}[Self cognitive skill {[}+{]}]                                &                                       &                                              \\
\tabitem Practitioner's predicting ability                                                                                                   &                                                                              &                                       &                                              \\
\tabitem Practitioner's timely working ability                                                                                               &                                                                              &                                       &                                              \\ \cmidrule(r){1-3}
\begin{tabular}[c]{@{}l@{}}\tabitem Practitioner perceiving team lacks cohesion\end{tabular}                                        & \hspace{-1em}\rdelim\}{8.8}{*}[Social cognition {[}+{]}]                                    & \hspace{-1em}\rdelim\}{8.8}{*}[Social cognition]     &                                              \\
\tabitem Practitioner perceiving team lacks skills                                                                                                        &                                                                              &                                       &                                              \\
\tabitem Practitioner perceiving team has skills                                                                                                          &                                                                              &                                       &                                              \\
\begin{tabular}[c]{@{}l@{}}\tabitem  Practitioner perceiving the lack of \\ emotional intelligence (social awareness:empathy) \\of the manager\end{tabular}  &                                                                              &                                       &                                              \\
\begin{tabular}[c]{@{}l@{}}\tabitem Practitioner perceiving the lack of\\ emotional intelligence (social awareness:\\empathy) of the customer\end{tabular} &                                                                              &                                       &                                              \\ \bottomrule
\end{tabular}%
}
\end{table*}

\textit{At the end of combined qualitative data analysis: }The strongest categories identified through our combined qualitative data analysis are as below.

\begin{tcolorbox}[colback=gray!10, boxrule=0pt,frame hidden]
\textbf{Categories:} Stimuli, Emotions, Emotion dynamics, Distinct events, Temporal matters
\end{tcolorbox}

We present these categories in detail in Section \ref{sec:findings}. We further found some emerging relationships as expected by the end of the STGT basic analysis, and we share these in Section \ref{sec:imp_research} in the form of an emerging theoretical model.

\section{Findings}
\label{sec:findings}
We found evidence of emotional responses to RCs, emotion dynamics in project and RC handling life cycles through our analysis. Further, we found the distinct events at which emotions are triggered belong to the project (project milestones) and RC handling life cycle (RC stages). We found some temporal matters which regulate the emotions felt, and the stimuli associated with emotion dynamics and regulation. 

When both high and low pleasurable emotions are found at a given event, we mention it by the term \textit{Mix}. In mix cases, we indicate the dominating emotion sub-scale within brackets. That is the sub-scale for which we found the most number of emotions. 

\begin{table}[t]
\caption{Emotional Responses to Requirements Changes}
\label{tab:all_emotions}
\resizebox{\columnwidth}{!}{%
\begin{tabular}{@{}lccccc@{}}
\toprule
\textbf{Emotion} & \multicolumn{1}{l}{\textbf{Never}}                     & \multicolumn{1}{l}{\textbf{Rarely}}                    & \multicolumn{1}{l}{\textbf{Sometimes}}                 & \multicolumn{1}{l}{\textbf{Quite Often}}               & \multicolumn{1}{l}{\textbf{Extremely Often}} \\ \midrule
\multicolumn{6}{l}{\textbf{High Pleasurable – High Arousal Emotions}}                                                                                                                                                                                                                              \\ \midrule
Ecstatic         & \cellcolor[HTML]{C0C0C0}22.39\%                        & \cellcolor[HTML]{9B9B9B}26.87\%                        & \cellcolor[HTML]{656565}{\color[HTML]{FFFFFF} 29.85\%} & \cellcolor[HTML]{C0C0C0}14.93\%                        & 5.97\%                                       \\
Energetic        & 5.47\%                                                 & \cellcolor[HTML]{C0C0C0}15.92\%                        & \cellcolor[HTML]{9B9B9B}27.86\%                        & \cellcolor[HTML]{656565}{\color[HTML]{FFFFFF} 32.84\%} & \cellcolor[HTML]{C0C0C0}17.91\%              \\
Enthusiastic     & 7.46\%                                                 & \cellcolor[HTML]{C0C0C0}11.94\%                        & \cellcolor[HTML]{9B9B9B}{\color[HTML]{333333} 28.86\%} & \cellcolor[HTML]{656565}{\color[HTML]{FFFFFF} 34.33\%} & \cellcolor[HTML]{C0C0C0}17.41\%              \\
Excited          & 8.46\%                                                 & \cellcolor[HTML]{C0C0C0}16.42\%                        & \cellcolor[HTML]{656565}{\color[HTML]{FFFFFF} 29.85\%} & \cellcolor[HTML]{9B9B9B}{\color[HTML]{333333} 24.88\%} & \cellcolor[HTML]{C0C0C0}20.40\%              \\
Inspired         & 9.45\%                                                 & \cellcolor[HTML]{C0C0C0}15.42\%                        & \cellcolor[HTML]{9B9B9B}23.88\%                        & \cellcolor[HTML]{656565}{\color[HTML]{FFFFFF} 30.35\%} & \cellcolor[HTML]{C0C0C0}20.90\%              \\ \midrule
\multicolumn{6}{l}{\textbf{High Pleasurable – Low Arousal Emotions}}                                                                                                                                                                                                                                \\ \midrule
At-ease          & \cellcolor[HTML]{EFEFEF}2.99\%                         & \cellcolor[HTML]{C0C0C0}23.88\%                        & \cellcolor[HTML]{656565}{\color[HTML]{FFFFFF} 39.30\%} & \cellcolor[HTML]{9B9B9B}26.37\%                        & 7.46\%                                       \\
Calm             & 2.49\%                                                 & \cellcolor[HTML]{C0C0C0}17.91\%                        & \cellcolor[HTML]{9B9B9B}29.35\%                        & \cellcolor[HTML]{656565}{\color[HTML]{FFFFFF} 40.30\%} & \cellcolor[HTML]{EFEFEF}9.95\%               \\
Content          & 6.97\%                                                 & \cellcolor[HTML]{C0C0C0}21.89\%                        & \cellcolor[HTML]{656565}{\color[HTML]{FFFFFF} 31.34\%} & \cellcolor[HTML]{9B9B9B}29.85\%                        & \cellcolor[HTML]{EFEFEF}9.95\%               \\
Relaxed          & \cellcolor[HTML]{EFEFEF}10.95\%                        & \cellcolor[HTML]{C0C0C0}15.92\%                        & \cellcolor[HTML]{656565}{\color[HTML]{FFFFFF} 33.33\%} & \cellcolor[HTML]{9B9B9B}30.85\%                        & 8.96\%                                       \\
Satisfied        & 6.47\%                                                 & \cellcolor[HTML]{EFEFEF}14.93\%                        & \cellcolor[HTML]{9B9B9B}27.36\%                        & \cellcolor[HTML]{656565}{\color[HTML]{FFFFFF} 28.36\%} & \cellcolor[HTML]{C0C0C0}22.89\%              \\ \midrule
\multicolumn{6}{l}{\textbf{Low Pleasurable – High Arousal Emotions}}                                                                                                                                                                                                                                \\ \midrule
Angry            & \cellcolor[HTML]{C0C0C0}16.42\%                        & \cellcolor[HTML]{9B9B9B}31.84\%                        & \cellcolor[HTML]{656565}{\color[HTML]{FFFFFF} 40.80\%} & \cellcolor[HTML]{EFEFEF}9.45\%                         & 1.49\%                                       \\
Anxious          & \cellcolor[HTML]{C0C0C0}10.95\%                        & \cellcolor[HTML]{9B9B9B}24.38\%                        & \cellcolor[HTML]{656565}{\color[HTML]{FFFFFF} 34.83\%} & \cellcolor[HTML]{9B9B9B}24.38\%                        & 5.47\%                                       \\
Disgusted        & \cellcolor[HTML]{656565}{\color[HTML]{FFFFFF} 44.28\%} & \cellcolor[HTML]{9B9B9B}26.37\%                        & \cellcolor[HTML]{C0C0C0}15.92\%                        & \cellcolor[HTML]{EFEFEF}10.95\%                        & 2.49\%                                       \\
Frightened       & \cellcolor[HTML]{9B9B9B}27.36\%                        & \cellcolor[HTML]{656565}{\color[HTML]{FFFFFF} 31.84\%} & \cellcolor[HTML]{C0C0C0}25.87\%                        & \cellcolor[HTML]{EFEFEF}10.95\%                        & 3.98\%                                       \\
Furious          & \cellcolor[HTML]{656565}{\color[HTML]{FFFFFF} 30.85\%} & \cellcolor[HTML]{9B9B9B}29.35\%                        & \cellcolor[HTML]{C0C0C0}21.39\%                        & \cellcolor[HTML]{EFEFEF}13.43\%                        & 4.98\%                                       \\ \midrule
\multicolumn{6}{l}{\textbf{Low Pleasurable – Low Arousal Emotions}}                                                                                                                                                                                                                                 \\ \midrule
Bored            & \cellcolor[HTML]{656565}{\color[HTML]{FFFFFF} 31.34\%} & \cellcolor[HTML]{9B9B9B}26.37\%                        & \cellcolor[HTML]{C0C0C0}24.88\%                        & \cellcolor[HTML]{EFEFEF}14.93\%                        & 2.49\%                                       \\
Depressed        & \cellcolor[HTML]{9B9B9B}29.35\%                        & \cellcolor[HTML]{656565}{\color[HTML]{FFFFFF} 32.84\%} & \cellcolor[HTML]{C0C0C0}24.88\%                        & \cellcolor[HTML]{EFEFEF}10.45\%                        & 2.49\%                                       \\
Discouraged      & \cellcolor[HTML]{C0C0C0}24.38\%                        & \cellcolor[HTML]{656565}{\color[HTML]{FFFFFF} 31.34\%} & \cellcolor[HTML]{9B9B9B}28.86\%                        & \cellcolor[HTML]{EFEFEF}12.44\%                        & 2.99\%                                       \\
Fatigued         & \cellcolor[HTML]{EFEFEF}9.45\%                         & \cellcolor[HTML]{9B9B9B}28.86\%                        & \cellcolor[HTML]{656565}{\color[HTML]{FFFFFF} 34.33\%} & \cellcolor[HTML]{C0C0C0}20.40\%                        & 6.97\%                                       \\
Gloomy           & \cellcolor[HTML]{9B9B9B}24.88\%                        & \cellcolor[HTML]{656565}{\color[HTML]{FFFFFF} 32.34\%} & \cellcolor[HTML]{C0C0C0}22.39\%                        & \cellcolor[HTML]{EFEFEF}15.92\%                        & 4.48\%                                       \\ \bottomrule
\end{tabular}%
}
\end{table}
\subsection{Emotional Responses to Requirements Changes}
The results from quantitative analysis provides a high level view of the emotional responses to RCs as a whole. We found that all high pleasurable emotions are more commonly felt by the software practitioners than low pleasurable emotions when handling RCs. Among the 10 low pleasurable emotions, only  three low pleasurable emotions were found as commonly felt (\textit{anger, anxiety,} and \textit{fatigue}) when handling RCs. Table \ref{tab:all_emotions} summarises the feeling of particular emotion within their current/most recent project when handling RCs as reported by the participants. We considered a certain emotion was felt commonly if the highest number of responses were found for ``sometimes/quite often/extremely often'' options. Therefore, as indicated in the table, the participants felt: \textbf{High\textsuperscript{2} emotions: }\textit{enthusiastic} (34.33\% quite often), \textit{energetic} (32.84\% quite often), \textit{inspired} (30.35\% quite often), \textit{ecstatic} (29.85\% sometimes), \textit{excited} (29.85\% sometimes);
\textbf{High\textsuperscript{1} emotions: }\textit{calm} (40.30\% quite often), at ease (39.30\% sometimes), \textit{content} (31.34\% sometimes), \textit{relaxed} (33.33\% sometimes), \textit{satisfied} (28.36\% quite often);
\textbf{Low\textsuperscript{1} emotions: }\textit{angry} (40.80\% sometimes), \textit{anxious} (34.83\% sometimes); and
\textbf{Low\textsuperscript{2} emotion: }\textit{fatigued} (34.33\% sometimes)
when handling RCs.

\subsection{Stimuli Triggering Practitioners' Emotions}
\label{sec:stimuli}


While \textit{RCs} act as the central stimulus in our study, through our STGT analysis we found several stimuli across project milestones, RC stages, and temporal matters. By careful observation, we noticed that these  are not stand-alone stimuli, but properties of a set of associated stimuli that lead to the triggering of emotions of practitioners. These associated stimuli are stakeholders, including \textit{practitioners} themselves, their \textit{team}, their \textit{manager}, and their \textit{customer(s)}. 

\textbf{Central stimulus: RC.} The properties of an RC that 
enable the triggering of emotions include RC stability, its point of introduction, frequency of introduction, impact, definition, status, severity, and challenging nature. Among these, \textit{point of introduction} is the leading property that trigger the emotions of practitioners. 

\textbf{Associated stimuli: Stakeholders.} \textit{Practitioner as a stimulus:} The Practitioner’s individual conation, individual and social cognition make the practitioner themselves a stimulus. As practitioners perceive team dynamics, and emotional intelligence (EI) of their managers and customers, their social cognition is also associated with other human stimuli. \textit{Team as a stimulus:} Team dynamics, such as collective skills and cohesion, results in \textit{team} being the stimulus of emotional response to RCs. \textit{Manager as a stimulus:} Managers of the practitioners act as stimuli because of their EI and being the source and carriers of some RCs. \textit{Customer as a stimulus:} Similar to manager, customer’s EI and being the source and carriers of RCs make them a stimulus for the emotional responses to RCs of practitioners.


\subsection{Emotion Dynamics of Practitioners in Project Life Cycle}
\label{sec:PJ}

\begin{figure*}
        \centering
        \includegraphics[width=\textwidth]{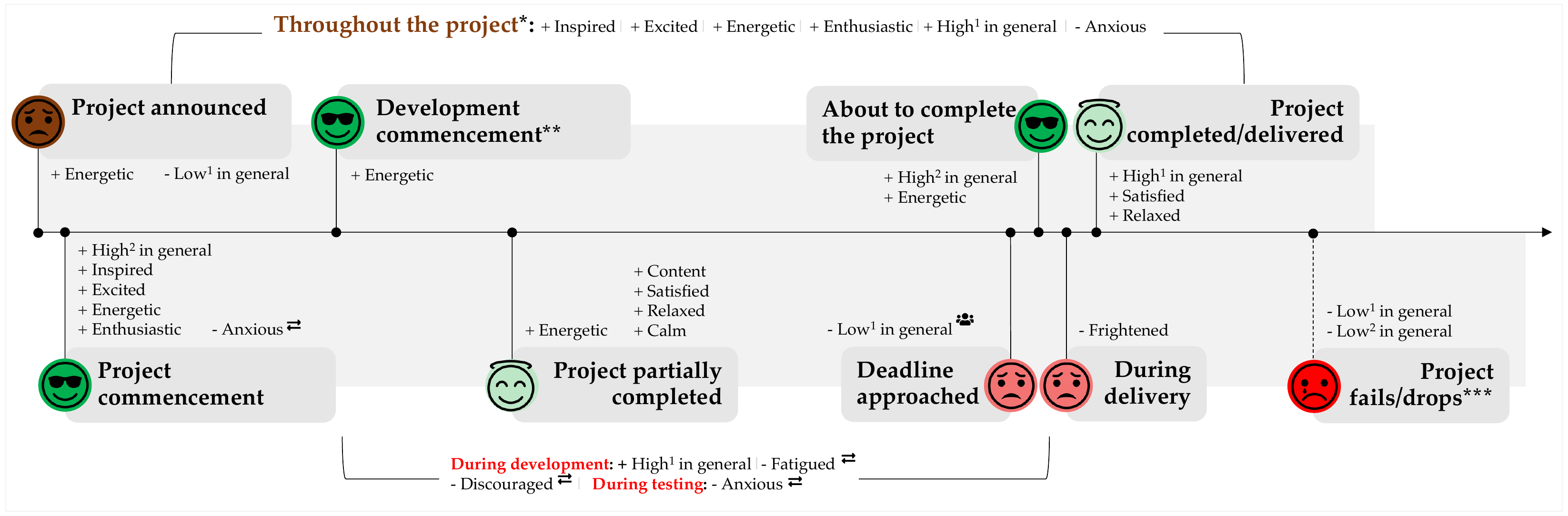}
        \caption{Emotion Dynamics of Practitioners in Project Life Cycle (*in general while specific emotions are triggered at specific milestones; **may overlap with the beginning of the project; ***may happen at any time; Emoji: Dominating emotion sub-scale; \includegraphics[scale=0.3]{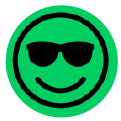}: High\textsuperscript{2}; \includegraphics[scale=0.3]{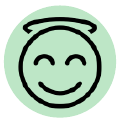}: High\textsuperscript{1}; \includegraphics[scale=0.3]{low1.pdf}: Low\textsuperscript{1}; \includegraphics[scale=0.3]{low2.pdf}: Low\textsuperscript{2};  \includegraphics[scale=0.6]{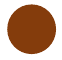}: both high/low emotions exist; Stimuli: \faExchange*: RC, \faBrain: Practitioner, \faUsers: Team, \faUserSecret: Manager, \faPeopleArrows: Customer)}
        \label{fig:project_emotions}
    \end{figure*}

We found critical milestones in a project life cycle where emotions of the practitioners are triggered. Fig. \ref{fig:project_emotions} demonstrates these milestones, the respective emotions triggered, and found stimuli at the milestones. While practitioners feel specific emotions at each milestone, we also found a mix of emotions occurs throughout the project -- low\textsuperscript{1} emotions during development, and in testing where \textit{anxiety} is common.

\subsubsection{Emotion Dynamics Pattern across Project Milestones}

The emotion dynamics pattern related to this case is given in Fig. \ref{project_pattern}. The emotions fluctuate in a \textbf{Mix (Low\textsuperscript{1}) $\rightarrow$ Mix (High\textsuperscript{2}) $\rightarrow$ High\textsuperscript{2} $\rightarrow$ Mix (High\textsuperscript{1}, Low\textsuperscript{1}) $\rightarrow$ Low\textsuperscript{1} $\rightarrow$ High\textsuperscript{2} 
$\rightarrow$ Low\textsuperscript{1} $\rightarrow$ High\textsuperscript{1}} manner. Low\textsuperscript{1} emotions were found \textit{at the beginning of the project}, then move towards high\textsuperscript{2} emotions. By the \textit{project's partial completion}, both high\textsuperscript{1} and low\textsuperscript{1} emotions were seen. Finally, by the \textit{end of the project}, when the \textit{deadline is approached} and \textit{during delivery}, low\textsuperscript{1} emotions can be seen.  When it is \textit{about to complete the project},  high\textsuperscript{2} emotions are abundant, whereas when the \textit{project is completed/delivered}, high\textsuperscript{1} emotions are plentiful. However, \textit{failure/drop of the project} could void this emotion dynamic pattern with low\textsuperscript{2} emotions. Therefore, \textit{project failure/drop} is the pattern cancelling milestone, which could happen at any time.

\subsubsection{Stimuli at Project Milestones}

\begin{figure}[b]%
    \centering
    \subfloat[\centering Project Life Cycle]{{\includegraphics[width=0.45\columnwidth]{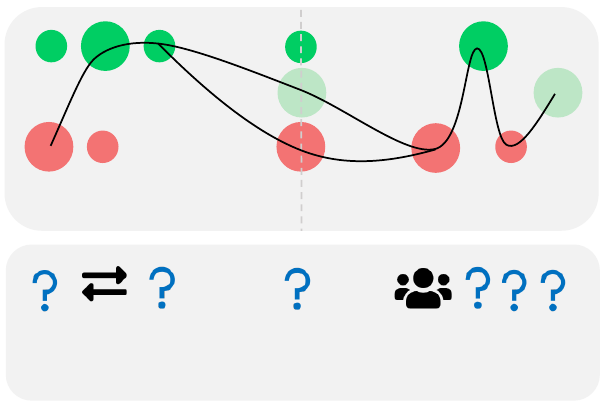}}
    \label{project_pattern}}%
    \qquad
    \subfloat[\centering RC handling Life Cycle]{{\includegraphics[width=0.45\columnwidth]{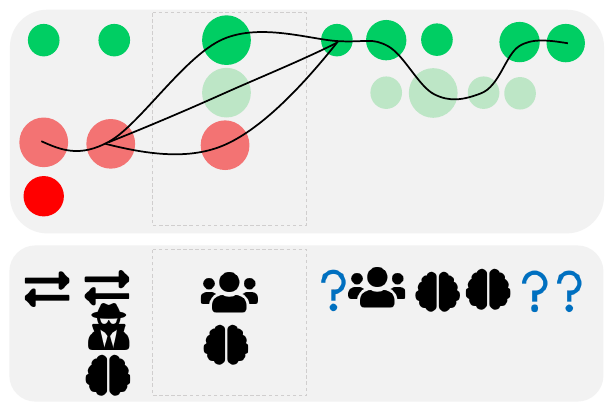}}
    \label{rc_pattern}}%
    \caption{Emotion Dynamics Patterns (1\textsuperscript{st} row: Dominating emotions; 2\textsuperscript{nd} row: Dominating stimuli; \includegraphics[scale=0.5]{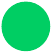}: high\textsuperscript{2}; \includegraphics[scale=0.5]{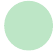}: high\textsuperscript{1}; \includegraphics[scale=0.5]{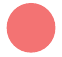}: low\textsuperscript{1}; \includegraphics[scale=0.5]{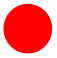}: low\textsuperscript{2}; Size of the circle: more the number of emotions, larger the circle; \faExchange*: RC; \faBrain: Practitioner; \faUsers: Team; \faUserSecret: Manager; \faPeopleArrows: Customer; 
    \includegraphics[scale=0.7]{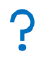}: Unknown)}%
    \label{fig:rc_pattern}%
\end{figure}

We did not find stimuli for all emotions at every project milestone. Hence, proving that triggering of emotions at the project milestones is not always due to stimuli, but only triggered because of the project milestone (which means stimuli decide the existence of emotion dynamics only occasionally). On the other hand, this is arguable that participants might not have revealed the associated stimuli. We leave this for researchers to investigate in the future. The same applies to emotions found at RC stages as well (following sub-section).
Below, within brackets next the project milestone, we present the found property of stimuli which contribute to the triggering of emotions at project milestone. Further, we present the dimensions, 
and evidencing quotes. 
The dominating stimulus at each milestone is shown in Fig. \ref{project_pattern}. 

\textbf{Project commencement (requirement stability):} 
Unstable requirements result in \textit{anxiety} at the commencement of the project: \textit{"I was anxious when we first started and the requirements were still not settled on." -- P54 [Tester].}

\textbf{During project development and testing (RC’s point of introduction):} During development, when RCs are introduced after design and implementation, practitioners are \textit{discouraged} and \textit{fatigued}: \textit{"I was fatigued in one such project (startup project), where almost every feature (3 out of 5 feature sets) had a change request in the midst of implementation phase. The most discouraging part is the time / phase when the requirement change happens. Most of the time it happens after the design and implementation phase." -- P31 [Agile Coach/Scrum Master, Developer];}


However, when an RC is introduced during testing, \textit{anxiety} is felt as reported by \textit{P159}: \textit{"During online trading system development after the development phases is completed and we are in testing phase of our product client called to roll back the whole segment of project which make me and my team anxious." -- P159 [Business Analyst].}

\textbf{Deadline approached (team dynamics):} As mentioned by \textit{P162}, as none of practitioners in the team could solve unresolved issues, low\textsuperscript{2} emotions occur: \textit{"The above emotions happened to me when the deadline of the project approached. At the last day of the release there some issues which couldn't be sorted out by anyone" -- P162 [Agile Coach/Scrum Master, Developer]}

\subsection{Emotion Dynamics of Practitioners in Requirements Change Handling Life Cycle}
\label{sec:RCS}

\begin{figure*}[b]
        \centering
        \includegraphics[width=\textwidth]{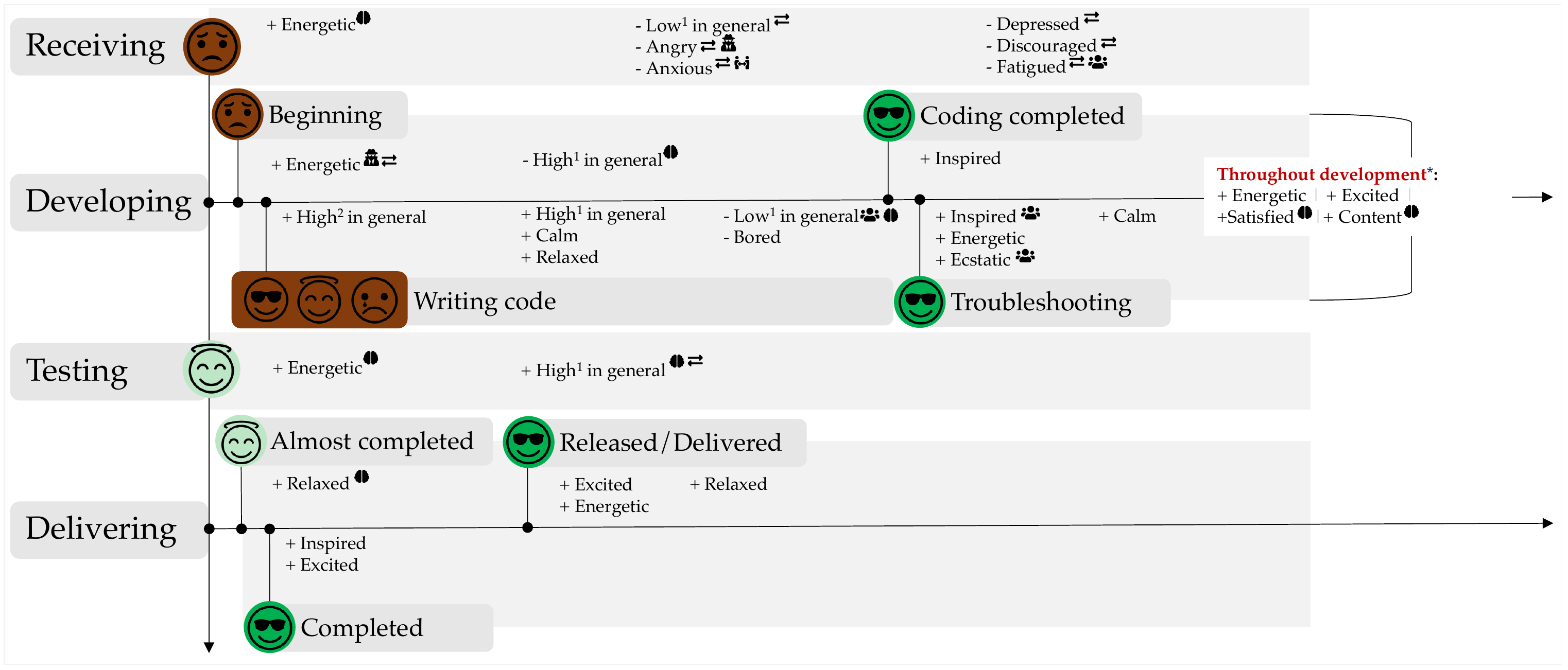}
        \caption{Emotion Dynamics of Practitioners in Requirements Change Handling Life Cycle (*in general while specific emotions are triggered at specific stages; Emoji: Dominating emotion sub-scale; \includegraphics[scale=0.3]{high2.pdf}: High\textsuperscript{2}; \includegraphics[scale=0.3]{high1.pdf}: High\textsuperscript{1}; \includegraphics[scale=0.3]{low1.pdf}: Low\textsuperscript{1}; \includegraphics[scale=0.3]{low2.pdf}: Low\textsuperscript{2}; \includegraphics[scale=0.6]{brown.pdf}: both high/low emotions exist; Stimuli: \faExchange*: RC, \faBrain: Practitioner, \faUsers: Team, \faUserSecret: Manager, \faPeopleArrows: Customer)}
        \label{fig:rc_emotions}
    \end{figure*}

We discovered additional stages of the RC handling life cycle to the stages we found in our preliminary study \cite{Madampe2020TowardsTeams}. Our initial research found the \textit{receiving, developing,} and \textit{delivering} stages where practitioners respond emotionally. In addition to that, through this study, we found the \textit{testing} stage, sub-stages of the developing stage (\textit{beginning, writing code, coding completed, troubleshooting}), sub-stages of delivering stage (\textit{almost completed, completed, released/delivered}). Fig \ref{fig:rc_emotions} gives a comprehensive illustration of these stages, corresponding emotions triggered at the stages, and found stimuli that trigger the emotions. Further, we found that practitioners feel a combination of high\textsuperscript{2} and high\textsuperscript{1} emotions \textit{throughout development} in general.

\subsubsection{Emotion Dynamics Pattern across Requirements Change Stages}

Fig. \ref{rc_pattern} shows the emotion dynamics pattern of the RC handling life cycle. The emotions fluctuate in a \textbf{Mix (Low\textsuperscript{1}) $\rightarrow$ Mix (Low\textsuperscript{1}) $\rightarrow$ Mix (High\textsuperscript{2}, High\textsuperscript{1}, Low\textsuperscript{1}) $\rightarrow$ High\textsuperscript{2} $\rightarrow$ High\textsuperscript{2} $\rightarrow$ High\textsuperscript{1} $\rightarrow$ High\textsuperscript{1} $\rightarrow$ High\textsuperscript{2} $\rightarrow$ High\textsuperscript{2}} way. Only 2 high\textsuperscript{2} emotions (\textit{energetic, excited}) were found at the \textit{receiving} stage, and the rest were low\textsuperscript{1} and low\textsuperscript{2} emotions. At the \textit{developing} stage, a mix of emotions was found where the domination of high\textsuperscript{2}, high\textsuperscript{1}, and low\textsuperscript{2} while \textit{writing code} was seen. However, when the \textit{coding was completed} and during \textit{troubleshooting}, high\textsuperscript{2} emotions were prominent. Then, by the \textit{testing} and \textit{delivering} stages, high\textsuperscript{1} and high\textsuperscript{2} emotions were seen.

\subsubsection{Stimuli at Requirements Change Stages}

Similar to project milestones, in this sub-section we present the 
stimulus property contributing to emotion triggering next to the RC stage where it was found, and quotes from participants. The dominating stimulus at each RC stage is shown in Fig. \ref{rc_pattern}.

\textbf{Receiving stage (RC’s point of introduction, impact of RC on other requirements, RC definition, RC type, carrier of RC, requirement changing extent, practitioners’ individual conation):}
The majority of the stimuli found at the receiving stage of an RC were related to the low\textsuperscript{1} and low\textsuperscript{2} emotions.
RC’s point of introduction -- such as receiving after development is significantly/fully completed, during testing and in the middle of the project, e.g.: \textit{``I would think we were done and then more changes would be asked as others reviewed. It left me depressed and fatigued at times.'' -- P191 [Business Analyst]};

Impact of RC on other requirements, where significant amount of inter-dependencies exist, is another stimuli which triggers the emotions of practitioners. Similarly, definition of RC, i.e., its initial learning, and 
type of RC such as improvements for better user experience, also trigger practitioners' emotions at this stage. Also, source and carrier of RC such as manager, and customers also trigger practitioners' emotions such as \textit{anxiety}, e.g.: \textit{``This is the 1st phase of every requirement change if it is the client who is initiating it. I feel anxious at the beginning but nothing else.'' -- P22 [Business Analyst];}


Practitioners’ individual conation -- where the practitioner is motivated to put much effort to work on the RC -- makes the practitioners \textit{energetic} when they receive the RC, e.g.: \textit{``Usually I feel more energetic when the work is assigned and there's a lot of work to do'' -- P148 [Developer, Tester]}.

\textbf{Development stage (Practitioners’ individual conation, individual cognition, team dynamics, RC’s code status):} Practitioners are \textit{energetic} in cases where they are motivated, e.g.: \textit{``I felt energetic at the start of the development as I haven't coded for a while.'' -- P9 [Developer]};

High\textsuperscript{2} emotions occur when the team is rapidly solving issues i.e., where positive team dynamics occur, e.g.: \textit{``As a crack team, bracing for any sort of change, especially changes in requirements, was like second nature to us. Whether planned or abrupt, we implemented any and all changes without needing to break much of a sweat. Sure, there were moments - quite rare, honestly - when we ran into some difficulties, but getting to troubleshoot and then rapidly resolve them left us all ecstatic and - if nothing else - greatly inspired.'' -- P101 [Manager]};

High\textsuperscript{1} emotions, such as \textit{satisfaction} and \textit{contentment}, are prominent where attention is sustained for a certain period of time, a cognitive skill, e.g.: \textit{``After a few solid hours of coding I did feel some satisfaction and content.'' -- P9 [Developer]};

Negative team cohesion, a team dynamic, such as when difficult for practitioners to agree, result in low\textsuperscript{1} emotions, e.g.: \textit{``I felt this emotion at the beginning because there were some obstacles with disagreements between the developers'' -- P181 [Developer, Manager, Product Owner, Tester];} 

When the code works well, the practitioners feel high\textsuperscript{2} emotions, e.g.: \textit{``I felt all this when the written codes works well.'' -- P65 [Tester].}

\textbf{Testing stage (Practitioners’ individual cognitive skills):}
Timely fixing of bugs and tests passing as expected trigger high\textsuperscript{1} emotions of the practitioners, e.g.: \textit{``energetic at every time that the testing was occurring as expected.'' -- P97 [Tester].}

\textbf{Delivering stage (Practitioners’ self-efficacy):}
We found that when practitioners' self-efficacy reported by \textit{P89}, where practitioners know the possibility of successfully delivering the RC, trigger a more relaxed frame of mind in them: \textit{``When I was closer to the end and all the most dangerous and boring parts had passed, I ended up relaxing because I knew I could deliver.'' -- P89 [Developer].}

\subsection{Emotion Dynamics Regulation by Temporal Matters}
\label{sec:TM}
    
\begin{figure*}
        \centering
        \includegraphics[width=\textwidth]{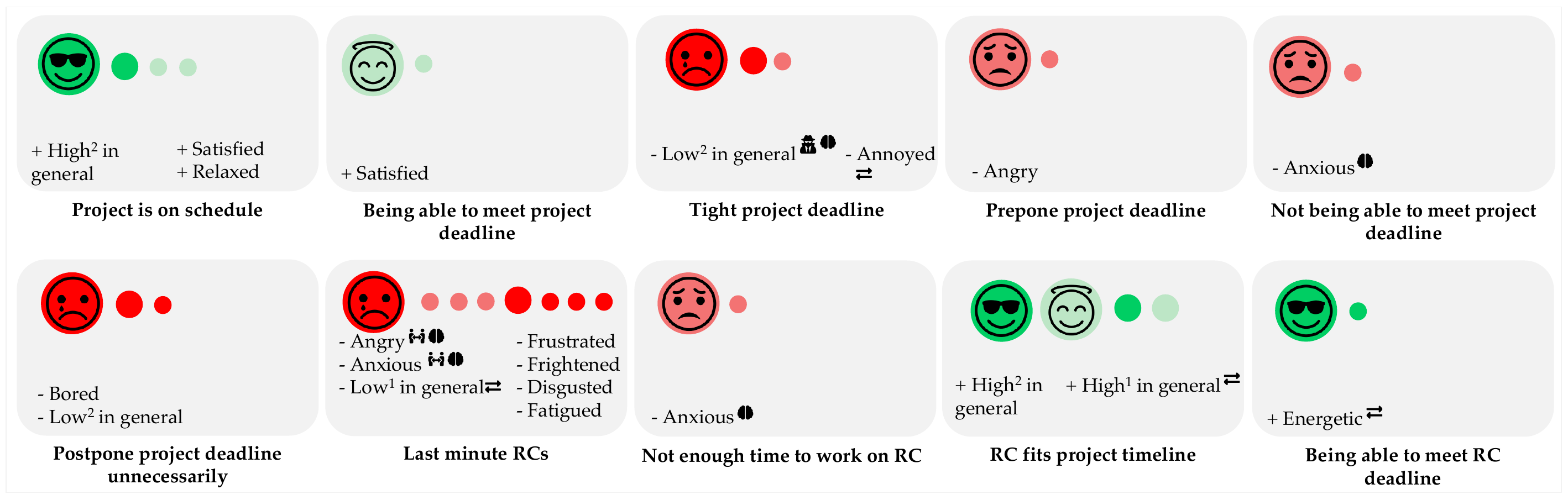}
        \caption{Temporal Matters which Regulate Emotional Responses (Emoji: Dominating emotion sub-scale; \includegraphics[scale=0.3]{high2.pdf}: High\textsuperscript{2}; \includegraphics[scale=0.3]{high1.pdf}: High\textsuperscript{1}; \includegraphics[scale=0.3]{low1.pdf}: Low\textsuperscript{1}; \includegraphics[scale=0.3]{low2.pdf}: Low\textsuperscript{2}; Stimuli: \faExchange*: RC, \faBrain: Practitioner, \faUsers: Team, \faUserSecret: Manager, \faPeopleArrows: Customer)}
        \label{fig:temporal_matters}
    \end{figure*}

We found that temporal matters can regulate practitioners' emotions. Fig. \ref{fig:temporal_matters} shows the key temporal matters and related emotional responses of the practitioners in our survey. 

\textbf{Temporal matters exclusive to the project:}
When a project is on schedule and practitioners are able to meet the project deadline, they feel high\textsuperscript{2} and high\textsuperscript{1} emotions, especially they are \textit{satisfied}, e.g.:
\textit{``When we finish a good project at a correct time it was satisfied at the moment. And it was the feeling of relief'' - P39 [Manager].}
But when the project has a tight deadline, and when the practitioners are not able to meet this deadline, practitioners feel low\textsuperscript{2} and low\textsuperscript{1} emotions, e.g.: \textit{``time was running out and we had a lot to do'' - P98 [Developer]}

If unnecessary deadline postponing happens, low\textsuperscript{1} and low\textsuperscript{2} emotions are felt by the practitioners, e.g.: \textit{``The project did not get sign-off at appropriate times which dragged the project on longer than it should have.'' - P183 [Business Analyst]}

\textbf{Temporal matters exclusive to RC handling life cycle:}
As reported by our participants, a widespread incident they face in their projects are ``last minute" RCs. Such a ``last minute'' could be closer to a set deadline. i.e., could be by the end of an iteration where a feature is anticipated to be released, or by the end of the project. We found that they feel low\textsuperscript{2} emotions due to such last minute RCs, e.g.: \textit{``A client wanted several significant changes made shortly before a deadline. I was very angry because they should've mentioned this earlier in development and having to make broad sweeping change so close to the deadline was very frustrating.'' - P138 [Developer]}

We also found that they feel low\textsuperscript{1} emotions, such as \textit{anxiety}, when they do not have enough time to work on the RC, e.g.: \textit{``I felt anxious because I was worried about whether or not I had enough time to make the change and complete the project on time'' - P150 [Developer, Tester]}

However, when they are able to meet the RC deadline, they feel high\textsuperscript{2} emotions, e.g.: \textit{``When they added a few usability enhancements due for the next day and I managed to set them in time for the deadline.'' - P180 [Tester]}

\subsubsection{Emotion Regulation of Project and RC Handling Life Cycles}
If an RC fits the project timeline, practitioners feel high\textsuperscript{2} and high\textsuperscript{1} emotions, thus making it the way to direct emotion regulation towards high pleasure, e.g.:
\textit{``When useful user-facing features are being added, when there is time to do so.'' -- P10 [Developer],
``When there's plenty of time for implementing new requirements (rarely) because there's no reason to be up'' -- P180 [Tester],
``If there is enough time, and the change is not a big modification in which I have to undone my work.'' -- P125 [Developer],
``This happens only when the changes are easy to implement and are not time-consuming. Some examples are the wordings on screens, minor layout changes, additional variations, or changes that can be implemented without major changes to the existing application design or program flow. Such changes are not a cause for concern and are usually easy to accommodate within the original estimates.'' -- P128 [Developer]}

\subsubsection{Stimuli during Temporal Matters}
The majority of stimuli during temporal matters we found were about the RC, especially if its \textbf{impact on scope} is minor, e.g.: \textit{``The project added a minor and easy to develop requirement which did not add significantly to the timeline or scope.'' - P183 [Business Analyst]} and more significant changes on scope, e.g.: \textit{``When the project scope changes and extended the project.'' - P170 [Business Analyst]}, \textbf{on other requirements}, e.g.: \textit{``When clients made last minute adjustments to planned feature'' - P59 [Developer]}, and on design/program flow.

We found that an \textbf{RC's challenging nature} is also a significant stimulus in temporal matters, especially when they are easy to implement, and not consuming time, e.g.: \textit{``This happens only when the changes are easy to implement and are not time consuming. Some examples are the wordings on screens, minor layout changes, additional variations, or changes that can be implemented without major changes to the existing application design or program flow. Such changes are not a cause for concern and are usually easy to accommodate within the original estimates.'' - P128 [Developer]}, and a stimulus related to \textbf{RC's delivery}, e.g.: \textit{``When they added a few usability enhancements due for the next day and I managed to set them in time for the deadline.'' -- P180 [Tester]} was found.

We also found stimuli related to \textbf{lack of EI of the customer}, and \textbf{social cognition of the practitioner}. The practitioners are able to perceive that the customer lacks EI (social awareness: empathy), e.g.: \textit{``Late in the dev phase of a new work project, my client changed their requirements; specifically they wanted to change how their data is loaded into the application, changing it from an XLS file to a database table. I had to rebuild the file loader completely. It made me feel angry because I felt they do not understand my position.'' -- P150 [Developer, Tester],} \textbf{individual conation} where practitioners are demotivated due to lack of EI of customer and when they act as the source, and carrier of RC, e.g.:
\textit{``Well we had almost completed the given task suddenly client called and changed their idea to an whole new thing. That moment I felt that my hard work was just wasted I felt very anxious while deleting and creating a new one crystal orange.'' -- P142 [Developer]}, and \textbf{perceiving their own personality}, e.g.: \textit{``I felt anxious because I was worried about whether or not I had enough time to make the change and complete the project on time, otherwise it would just look bad on me regardless of the requirements being changed, and I am a perfectionist.'' -- P150 [Developer, Tester]}, and practitioners \textbf{social cognition} where they are perceiving their manager \textbf{lacks EI}, e.g.:
\textit{``I worked more and more time each day to try to successfully make the job, and new issues appeared every day that needed more and more time. We were all at same feelings but team leader didn"t hear his team feeling. He requested us to perform this project on at 'just in time'. I lost energy and I worked without any positive feedback. I lost my flow.'' -- P173 [Developer]
}

\section{Threats to Validity}
\label{sec:threats}


\textbf{External Validity:}
Equal geographic distribution of participants was not achieved. Almost half of the participants of our study were from North America, and therefore, generalizability of our findings is limited.
Similarly, participant counts across the genders was not equal. Majority of participants were male, rest were female except a single gender-diverse participant.  As emotions are bounded to the biological nature of humans, including the gender, we see this as a threat to validity.
However, in both of the above-mentioned cases, we tried our best to recruit representative participants.
This study was conducted during Covid-19 global pandemic. Mental health was a persistently talked and researched topic during this time. As emotions heavily impact the mental well-being, the results of this study may have been threatened by the pandemic situation. A study on work of software engineers during Covid-19 also found the impact of it on the mental well-being \cite{Ralph2020PandemicHelp}. However, this situation was an uncontrollable one, therefore stays as a threat to the validity of the findings.

\textbf{Internal Validity:}
The first author analysed all data and the emerging findings were presented to the second and third authors during fortnight meetings where the findings were discussed. In addition, all authors went through the codes individually as a final check before drafting the paper. We also consulted a psychology expert to ensure the correctness of terminology use, and analysis of data.

\textbf{Construct Validity:}
Given the Covid-19 pandemic situation, we were not able to conduct any experience sampling or observations to collect emotional responses shown at the exact moment, as we planned. 
Therefore, we decided to carry out an online survey allowing participants to self-report their emotions. However, we tried to mitigate this threat by asking the participants to share their experiences considering their current/most recent project. 
Our findings rely on JAWS. JAWS is used to assess emotional reactions of people over the past 30 days. However, in our case, it is impossible to assume that project and RC handling life cycles were limited to a period of 30 days. Therefore, this remains as a threat to validity. Additionally, the number of emotions vary across the different emotion scales. Therefore, the analysis may differ if other emotion scales are used. In addition, we assumed that emotions listed in the scale are understandable and interpreted in the same way by the participants. 
Moreover, as participants may be reluctant to report their negative emotions \cite{Wrobel2013}, this is a possible threat. The recommendations we  presented assume that high pleasurable emotions impact handling of RCs positively and low pleasurable emotions impact handling of RCs negatively in general. However, certain emotions such as \textit{anger} might sometimes make software teams more productive \cite{Wrobel2013}. 

\section{Discussion}
\label{sec:discussion}
\subsection{Implications for Practitioners}
\textbf{Practitioners are not always \textit{pleased} with RCs.} 
The majority of our survey participants practice agile methods. The Agile Manifesto claims that agile approaches allow practitioners to respond better to changes without necessarily needing to follow a plan \cite{Beck2001ManifestoDevelopment}. Practitioners do respond to RCs, and the manifesto does not talk about whether the practitioners should respond positively or not i.e., the Agile Manifesto does not say whether the practitioners should be \textit{pleased} with all of their RCs or not. Even though the quantitative analysis gives a high level sense that practitioners feel high pleasurable emotions when handling RCs, according to our findings from in-depth qualitative analysis show that practitioners are not always pleased with RCs and show a mix of emotions throughout the RC stages. However, as agile is known to be human-centric, our findings around emotions, which play a major role in human mind, do raise the question, ``is agile really that human-centric?''

\textbf{The impracticality of welcoming changing requirements even late in development.} 
Even though the agile manifesto encourages welcoming changing requirements, even late in development, it is notably clear from our findings that last minute RCs trigger low\textsuperscript{1} and low\textsuperscript{2} emotions of  practitioners. By the end of the project, or even by the end of an iteration, if closer to a deadline, the above-mentioned emotions are felt when an RC is introduced. Therefore, \textit{positively} welcoming changing requirements even late in development is often impractical. 

\textbf{The agile principle, ``welcoming changing requirements even late in development" voids the principle of ``giving the software team the environment and support they need''.} Emotions have direct links to cognition, productivity, and decision-making, essential elements in developing software. According to our findings, it is evident that the stimuli occasionally decide the existence of emotion dynamics along with the distinct events. In other words, these stimuli and events have the ability to hinder the arousal of high pleasurable emotions in practitioners. As we found that last minute RCs are common,  and  as low\textsuperscript{1} and low\textsuperscript{2} emotions are triggered in practitioners when last minute RCs are introduced, it is clear that the environment and support in terms of emotions may not be being supplied to practitioners in SE team contexts. Therefore, in reality, positively welcoming changing requirements and giving the software team, the environment and support they need in terms of emotions is not always possible. 

\textbf{Practitioners having emotional intelligence, but their managers, and customers lack.} EI has four aspects, namely, self-awareness (aware of own emotions), self-management (manage own emotions), social-awareness (aware of others’ emotions), relationship management (build relationships with the use of other three aspects). \textit{Self-awareness:} From our results, it is clear that, practitioners are generally aware of their own emotions as they explained ``when they emotionally respond” by incorporating emotions clearly. \textit{Self-management:} Practitioners did not mention how they manage their emotions at the distinct events we have given in this paper. 
\textit{Social-awareness:} Our study shows that practitioners recognise when customers, and their managers, lack EI; which is a sign of practitioners’ capability of being socially aware of others. However, we could not find practitioners recognising specific emotions of others at the distinct events.  \textit{Relationship management:} Even though our results show that the practitioners identify the relationship statuses such as disagreements among developers at a distinct event, we did not find how they improve the relationship to build a better relationship. Even though practitioners' EI is impressive, the same EI does not seem to be perceived in their managers and customers. Them  not being empathetic (social-awareness) arouses low pleasurable emotions of the practitioners.

\begin{figure*}
        \centering
        \includegraphics[width=\textwidth]{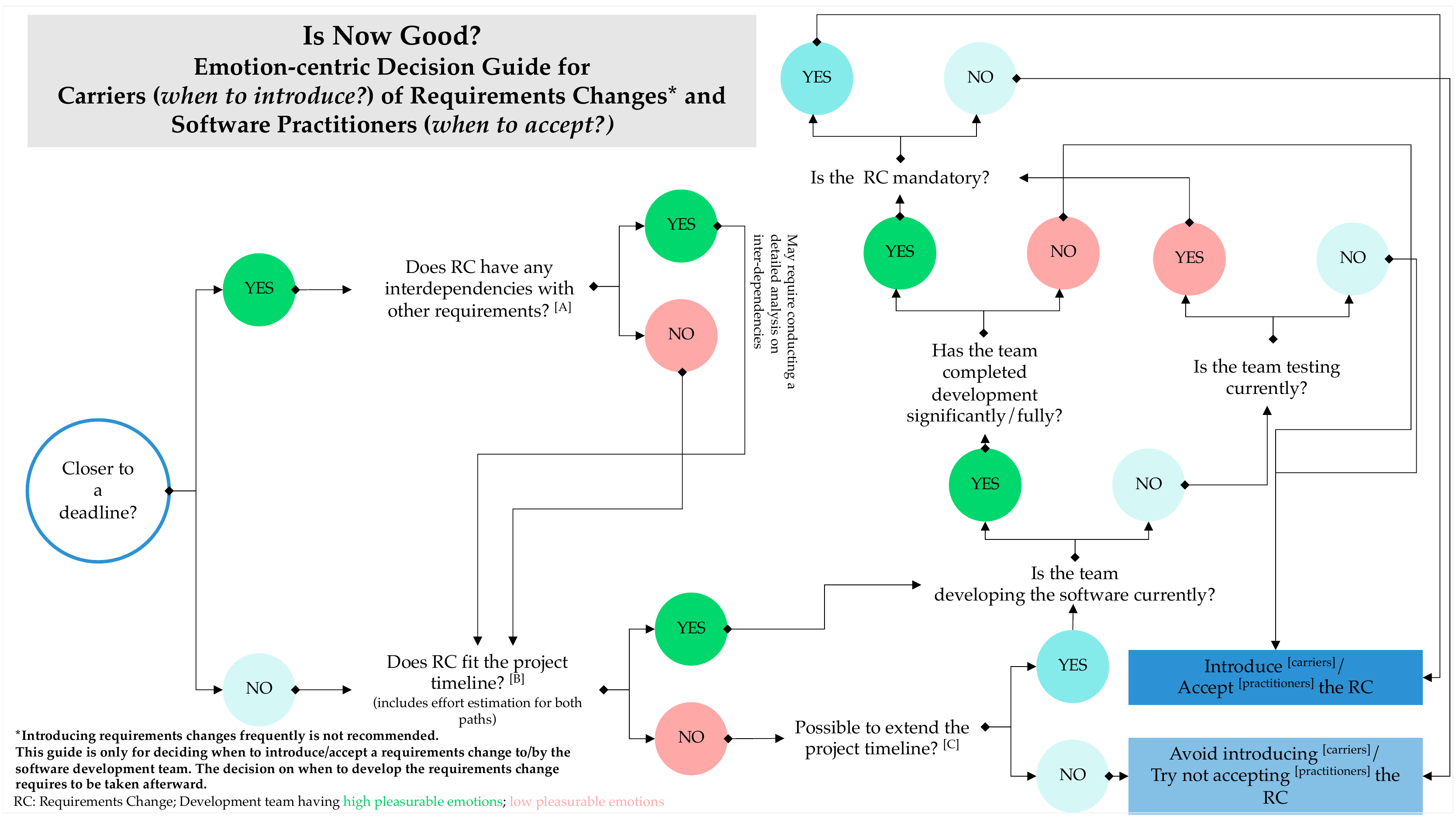}
        \caption{Dual-purpose Guide: Emotion-centric Decision Guide for Carriers of Requirements Changes and Practitioners}
        \label{fig:decision_guide}
    \end{figure*}

\textbf{Emotion-centric Decision Guide.} 
We present an emotion-centric decision guide that we created using the above implications in Fig. \ref{fig:decision_guide}. This dual-purpose guide caters for the carriers of RCs by suggesting to them when to decide when to introduce an RC, and for practitioners to decide when to accept an RC by being considerate about their emotions. The checkpoints [A], [B], and [C] require the carriers of RCs and practitioners to work collectively. 

\textbf{Recommendation 1. Practitioners monitoring their own emotions and managers monitoring team's emotions is necessary}. \textit{Anxiety} is common among practitioners. Monitoring emotions help improve their EI through self-awareness, and eventually help take actions (self-management) to maintain positive emotional well-being, thus providing the support the practitioners need in terms of emotions. Emotion tracking and monitoring tools exist. For instance, \textit{Emotimonitor} \cite{El-Migid2021Emotimonitor:Teams}, the tool we developed for agile teams, could be used to track and monitor self and team emotions. 

\textbf{Recommendation 2. Customers and managers should improve their emotional intelligence.} Customers and managers lacking EI impact the emotional well-being of practitioners. As practitioners have EI, they do know when their managers, and customers lack EI. To have a better relationship with the practitioners, customers and managers are required to improve their EI. Since we found lack of empathy is identified by the practitioners, customers and managers, empathy is specifically needed to be improved and prioritised. For example, in this article \cite{ProductiveEmpathy}, the author explains how prioritising empathy could have made a conversation more productive. Customers and managers being concerned and ``human'' requires to be prioritised for the betterment of the emotional well-being of the practitioners in the team.

\textbf{Recommendation 3. Being considerate of when to introduce/accept RCs is necessary.} In relation to the previous recommendation, the point of RC introduction plays a significant role in the emotional responses of practitioners. Avoiding or reducing last minute RCs may bypass the feeling of low\textsuperscript{1} and low\textsuperscript{2} emotions of the practitioners.

\textbf{Recommendation 4. When possible, extending the project timeline where necessary may help maintain high\textsuperscript{2} and high\textsuperscript{1} emotions of the practitioners.} For example, when there is not enough time for the practitioners to work on the RC, they are \textit{anxious}. However, when the RC fits the project timeline, practitioners feel high\textsuperscript{2} and high\textsuperscript{1} emotions. Therefore, allocating enough time, which may result in an extension of the project timeline, may awaken the practitioners' high\textsuperscript{2} and high\textsuperscript{1} emotions. However, this recommendation is  applicable if and only if the deadline extension is possible with all other business factors.

\textbf{Recommendation 5. Do not exploit agile values and principles.} Finally, we urge the carriers of RCs and practitioners not to exploit the agile value ``responding to change over following a plan'' by entirely avoiding following a plan, and the principle ``welcoming changing requirements even late in development''. Following guides which are kind of lightweight plans may ultimately help maintain better practitioner emotional well-being. 

\subsection{Implications for Researchers}
\label{sec:imp_research}
    
\textbf{Similarities and differences in emotion dynamics of project and RC handling life cycles.}
The emotion dynamics pattern of the RC handling life cycle is different from the project life cycle. \textbf{Similarity:} Both have low\textsuperscript{1} emotions dominating at the beginning. \textbf{Differences:} In the RC handling life cycle, low\textsuperscript{2} emotions such as \textit{depressed, discouraged}, and \textit{fatigued} can be seen, which is not seen at the beginning of the project. In the middle of the project life cycle, only high\textsuperscript{1} and low\textsuperscript{1} emotions are prominent. In the RC handling life cycle, high\textsuperscript{2} emotions are also central; in the project life cycle, by the end, high\textsuperscript{2}, high\textsuperscript{1}, and low\textsuperscript{1} emotions can be seen, but in the RC handling life cycle, no low\textsuperscript{1} emotions are evident. Therefore, we can say that practitioners are much pleased by the end of the RC handling life cycle than that of the project life cycle. Note of course that the RC handling life cycle begins within the project life cycle.

\textbf{Specific stimuli dominate the emotion dynamics at project milestones and RC stages.} A \textbf{mix of stimuli} dominates project milestones. However, stimuli were not found for the feeling of emotions when the \textit{project is announced}, at the \textit{development commencement}, when the \textit{project is partially completed}, \textit{during delivery}, and when the \textit{project is completed/delivered}. In the RC handling life cycle, the \textbf{practitioner as a stimulus} dominates the majority of the stages and then the RC. This hints that even though RC acts as the central stimulus, the practitioner's individual conation, individual cognition, and social cognition play a larger part in triggering their own emotions. We anticipate to see research on how practitioners explore, utilise, and improve these aspects. Similar to the project, we did not find the dominating stimuli when \textit{coding is completed,} and when the \textit{RC is completed}, and \textit{released}. We encourage researchers to study the missing dominating stimuli at the above-mentioned project milestones and stages of the RC handling life cycle.

\textbf{Other stimuli exist.} Apart from RCs, and the rest of human stimuli we have presented, we found other stimuli such as some properties of the project, issue, and task that trigger emotions of practitioners at distinct events. However, the evidence is not dense enough to accept them as triggers. A few such scenarios are given below, and researchers may consider exploring these more in the future.
\textbf{Project partially completed (project stability, issue severity):} 
When the project is stable and when there are no major issues at partial completion of the project,  high\textsuperscript{1} emotions arise, e.g.: \textit{"Usually somewhere around mid project, if everything is on schedule and there haven't been any major glitches, then most of us, if not just me will feel content and satisfied and even relaxed. The schedule can be hectic at times but that doesn't mean it can't be approached in a calm and relaxed manner." -- P64 [Manager, Tester]}; 
\textbf{Project completed/delivered (project completion status):} We found that when the project was successfully delivered, \textit{satisfaction} occurs; 
\textbf{Throughout the project (project stability):} Project undergoing a stable implementation impact the high\textsuperscript{1} and high\textsuperscript{2} emotions, e.g.: \textit{ "Any time our team begins any project we all start out with enthusiasm, inspiration and a great deal of energy. Hopefully, those feelings stay with us through the whole project but they don't always depending on how the actual implementation goes" -- P64 [Manager, Tester]}.

\textbf{The emerging theoretical model.} At the end of STGT basic data analysis stage, we found emerging relationships among the categories. We noted these down and used them to aid further discovery and strengthening of the connections. The first author drew the emerging model on paper iteratively, and shared and discussed it with the other two authors until the final model was formed. The emerging relationships identified among categories are given in bold uppercase text below.

\begin{tcolorbox}[colback=gray!10, boxrule=0pt,frame hidden]
\textbf{Emerging relationships among categories:} \\
Emotion dynamics \textbf{IS THE FLUCTUATION} of emotions over time; Stimuli \textbf{TRIGGER} emotions; Distinct events \textbf{LEAD TO} emotion dynamics; Temporal matters \textbf{COMMON TO} distinct events \textbf{REGULATE} emotion dynamics
\end{tcolorbox}

\begin{figure}[]
    \centering
    \includegraphics[width=\columnwidth]{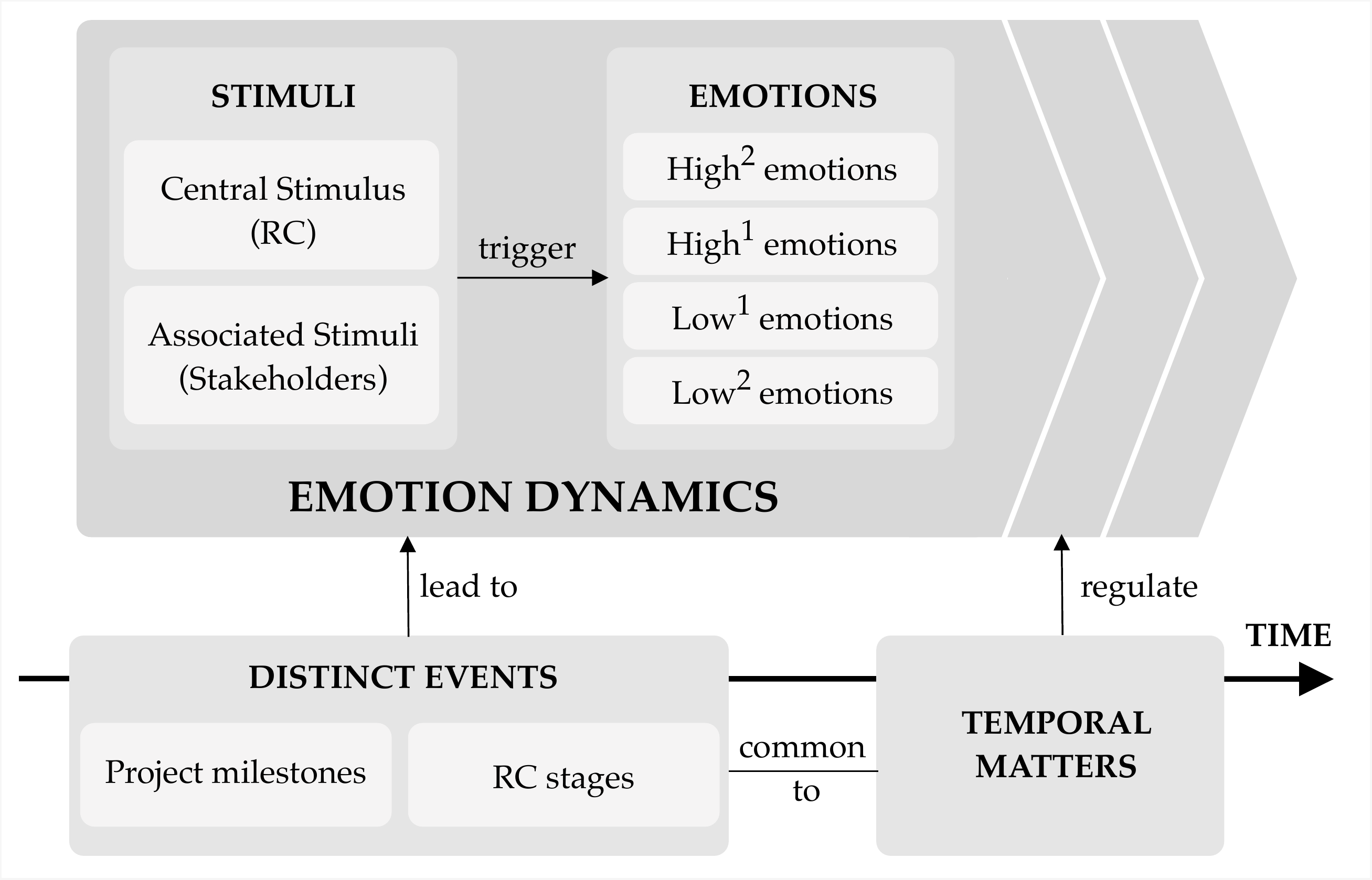}
    \caption{Emerging Theoretical Model: Emotion Dynamics of Software Team Contexts}
    \label{fig:model}
\end{figure}

We suggest future work on this, where the researchers could either choose the emergent or structured mode of STGT to develop a full theory. For instance, if the researchers choose the structured mode, the next steps include structured data collection, structured data analysis, advanced memoing, and finally theoretical integration. The model representing the emerging relationships is given in Fig. \ref{fig:model}. The core category at this stage is:

\begin{tcolorbox}[colback=gray!10, boxrule=0pt,frame hidden]
\textbf{Emerging core category (central phenomenon):} Emotion dynamics in software team contexts
\end{tcolorbox}

\textit{STGT Method Application Evaluation:} 
STGT outlines criteria for evaluating the application of the method. As this paper does not propose a mature theory, we evaluated our model against the criteria for non-mature theories (Credibility and Rigor).
\textbf{Credibility: }We have provided details in Section \ref{fig:mtd} on how participants were recruited (social media and AMT), the applied an initial sampling method (random sampling followed by purposive sampling), how iterative and interleaved data collection and analysis ensured (the study is not limited to this paper, therefore the iterative data collection and analysis was quantitative data-biased), and how memos written and used (diagrams). \textbf{Rigour:} In Section \ref{fig:mtd}, we have provided examples of our basic coding (how raw data was analysed to produce codes, subcategories, and categories), embedded sanitised evidence (quotes from throughout out the paper), and evidence of theory development (the emerging theory).

\section{Related Work}
\label{sec:RW}

\subsection{Emotions During/Post Software Development}

\textbf{Exploring Emotions:}
In their study, Graziotin et al. \cite{Graziotin2014SoftwarePerformance} summarise existing psychology and SE  studies to demonstrate practitioners calling for more psychology-based SE studies. Pletea et al. \cite{Pletea2014SecurityGitHub} conducted sentiment analysis of discussions and comments of commits and pull requests. They found that security related discussions on GitHub contain more negative emotions than other discussions.

The proposal paper \cite{Fountaine2017EmotionalMeasurement} on exploring emotions of software developers while working suggested that emotional awareness increases developer's progress by mitigating negative emotions. 
\cite{Kolakowska2013EmotionEngineering} is also a proposal paper on  multimodal emotion recognition of software developers while working. Application scenarios in software development and testing processes using multimodal emotion recognition (vision, sound, text, physiological signals). 
Novielli et al.'s proposal paper \cite{Novielli2014TowardsOverflow} on exploring emotion in questions and answers in Stack Overflow argues that emotions of a technical question impacts the probability of obtaining satisfying answers. They also provide an emotion annotated dataset \cite{Novielli2018AOverflow} of questions, answers, and comments from Stack Overflow which can be used in developing predictive models. 

Yang et al. \cite{Yang2017AnalyzingProjects} improved Naïve Bayes Multinomial algorithm for emotion analysis. 
Werder and Brinkkemper \cite{Werder2018MEME:Github} presented a tool for extracting emotions from GitHub comments. They used Ekman's and Davidson's emotional framework in their tool. Murgia et al. \cite{Murgia2018AnSystems} also presented a tool for extracting emotions from issue comments. They found that words such as \textit{thanks} and \textit{sorry} are emotion-driven. They found their machine learning classifiers for \textit{love} and \textit{joy} have an overlap between them. Guzman's study \cite{Guzman2013}  includes general and detailed views of topics and emotions expressed in software project collaboration artefacts. It suggested how topic modelling can be applied to extract sentiment score and topics in the text.  
Neupane et al.'s work \cite{Neupane2019EmoD:Development} is on automatic extraction of communication records of software development teams. Identification of emotions and their intensity, and modelling them as time series data, provides end to end support on data collection, modelling, storage, analysis, and presentation of emotions. Similarly a study on automatic extraction of emotions from issue comments \cite{Murgia2014DoArtifacts}  found that issue reports express emotions and emotions such as \textit{love, joy,} and \textit{sadness} can be automatically mined. Parrot's emotion framework was used in their study. 

\textbf{Emotions vs Productivity:}
Through their study \cite{Wrobel2013} focusing on the impact of emotions on productivity during software development, Wrobel found that frustration is felt most often. This lowers productivity, while anger increases productivity, enthusiasm increases productivity, and emotions transit from frustration $\rightarrow$ anger $\rightarrow$ contentment $\rightarrow$ enthusiasm. 
Wrobel also proposed participant observation can be used to conduct experiments on understanding emotions in SE teams \cite{Wrobel2016TowardsTeams}. This can be used as an assessment tool to measure valence and arousal of emotions, and productivity of SE teams. Crawford et al.'s work \cite{Crawford2014} on the relevance of emotions in software development noted the need of conducting emotion based research in software contexts. Giradi et al.'s \cite{Girardi2021EmotionsWorkplace} provide evidence on connection between emotions and productivity. 
They also present a taxonomy of triggers which includes collaboration, and self perception which we also found through our study.

\textbf{Emotions vs Progress:}
Muller and Fritz's work \cite{Muller2015} on relationship between emotions and progress of software developers concluded that emotions have direct potential impact on developer's work and productivity. They also found that the reasons for increase in emotions/progress: localising relevant code, better understanding of parts of the code, clear next steps, writing code, and having new ideas. As the reasons for decreases in emotions/progress, they found difficulty in understanding how parts of the code/API work, difficulty in localising relevant code, not being sure about next steps, realising that hypothesis on how code works is wrong, and missing/insufficient documentation. 
Girardi et al.'s work \cite{Girardi2018SensingExperiment} proposes replicating a previous study to determine the extent to which biometric sensors can be used to automatically detect emotions during software development to find the relationship between emotions and progress.

\textbf{Emotions vs Practices:}
In their study \cite{Colomo-Palacios2019EmotionsCoding}, Colomo-Palacious et al. 
compare emotions felt while presenting and coding. They found that anxiety and nervousness are felt when presenting and satisfaction and enjoyment are felt when coding.

\textbf{Emotions vs Problem Solving:}
Graziotin et al.'s work \cite{Graziotin2014SoftwarePerformance} on the relationship between affective states, creativity, and problem-solving skills demonstrated the need for more psychology-based SE studies.

\textbf{Affective States vs Software Metrics:}
Kuutila et al.'s study \cite{Kuutila2018UsingWell-being} on identifying links between software developers' affective states and work well-being, 
and found a negative link exists between hurry and number of commits, and a negative relationship exists between social interaction and hindered work well-being.

\subsection{Emotions During Requirements Engineering}

\textbf{Pleasure and Arousal of Emotions vs State of Requirements:}
Colomo-Palacious et al.'s study \cite{Colomo-Palacios2011UsingEngineering} on comparison of pleasure and arousal in final and non-final requirements, stated that pleasure felt is high in final requirements and arousal is low in final requirements. 

\textbf{Emotions vs Handling Requirements Changes:}
Previously, we \cite{Madampe2020TowardsTeams} presented distinct phases of RCs, a summary of emotions, experienced, and variation in emotions and sentiment polarity when \textit{receiving}, \textit{developing}, and \textit{delivering} RCs. We 
found that positive and neutral emotions follow high $\rightarrow$ low $\rightarrow$ high pattern whereas negative emotions are higher when RCs are received and moves towards positive as teams move to developing, and delivering RCs. In this paper, we present stages and emotions in a finer level.


\section{Conclusion}
\label{sec:conclusion}
In this paper, we present emotional responses to requirements changes, and \textit{ emotion dynamics}: how emotions of software practitioners fluctuate over time in their project when they are handling requirements changes. We found specific project milestones in the project life cycle and stages in the requirements change handling life cycle where practitioners emotions are triggered. We found that these emotions are not only tied to the project milestones and requirements change stages, but also to the \textit{stimuli} of requirements change, practitioner, team, manager, and customer. In addition, we discovered that regulation of emotions is possible through \textit{temporal matters}, which are universal to the project and to the requirements change handling life cycles. We conclude that practitioners are not always pleased with requirements changes, and that positively welcoming changing requirements  in late development is impractical, hence violating the emotional support provided to the practitioners working in the software development project. We propose a dual-purpose emotion-centric decision guide for the carriers of the requirements changes (customers and internal stakeholders who liaise with customers) and practitioners to decide when to introduce/accept a requirements change to/by the team. We also provide recommendations for practitioners to follow and directions for researchers to explore this area further in the future.
    
\section*{Acknowledgments}
This work is supported by a Monash Faculty of IT scholarship. Grundy is supported by ARC Laureate Fellowship FL190100035. Also, our sincere gratitude goes to Dr William Bingley for providing invaluable feedback for this work, and all the participants who took part in this study.

\bibliographystyle{ieeetr}
\bibliography{main}
\vspace{-6cm}

\end{document}